\documentclass[doublecol]{epl2}
\usepackage{graphicx}
\usepackage{amsfonts}		% open math symbols
\usepackage{amsmath}		% open math symbols
\usepackage{amssymb}		% open math symbols
\usepackage{color}
\usepackage{multirow}

\title{How soon after a zero-temperature quench  is the fate of the Ising model sealed?}
\author{T.~Blanchard\inst{1,2} \and F. Corberi\inst{3} \and L.~F.~Cugliandolo\inst{1,2} \and M.~Picco\inst{1,2}}
\shortauthor{T.~Blanchard, F. Corberi, L.~F.~Cugliandolo and M.~Picco}

\institute{
\inst{1}Sorbonne Universit\'es, UPMC Univ Paris 06, UMR 7589, LPTHE, F-75005, Paris, France\\
\inst{2}CNRS, UMR 7589, LPTHE, F-75005, Paris, France\\
\inst{3} Dipartimento di Fisica ``E. R. Caianiello'', Universit\`a  di Salerno,
via Ponte don Melillo, 84084 Fisciano (SA), Italia
}


\date{\today}

\abstract{
We study the transient between a fully disordered initial condition and a percolating structure in the 
low-temperature non-conserved order parameter dynamics of the bi-dimensional Ising model.
We show that a stable structure of spanning clusters establishes at a time $t_p \simeq L^{\alpha_p}$.
Our numerical results yield $\alpha_p=0.50(2)$ for
the square and kagome, $\alpha_p=0.33(2)$ for the triangular and $\alpha_p=0.38(5)$ for the 
bowtie-a lattices. We generalise the dynamic scaling hypothesis to take into account  this new
time-scale. We discuss the implications of these results 
for other non-equilibrium processes.
}

\begin{document}
\maketitle

Phase ordering kinetics is the process whereby an open classical system locally orders in its 
equilibrium states. This phenomenon occurs when a macroscopic system is taken across a second order phase 
transition by changing (slowly or abruptly) one of its parameters or the environmental conditions.
Theoretic and numeric approaches to phase ordering kinetics focus on the scaling regime, in which 
structures with {\it typical} linear size $\xi(t)$ satisfying $\xi_{micro} \ll \xi(t) \ll L$, with $\xi_{micro}$ a microscopic lenght-scale associated
to the lattice spacing and $L$ the 
linear size of the system, have established. The mechanism controlling the local growth of the equilibrium phases
can in general be identified, and allows one to deduce the time-dependence in $\xi(t)$~\cite{Bray94}.
The stochastic evolution of the ferromagnetic Ising model (IM) quenched below $T_c$ is a textbook example of coarsening
phenomenology, with $\xi(t)\sim t^{1/z}$ and $z=2$ for non-conserved order-parameter dynamics~\cite{Bray94}.
The role played by the initial conditions and the pre-asymptotic dynamics leading to the 
scaling regime, and {\it atypical} ordered spatial regions, in this and other models have not been studied in detail yet. 

Evidence for percolation~\cite{percolation} influencing the scaling regime and asymptotic states reached 
by the $2d$ square-lattice IM after a quench from high to low temperature were given by two groups. In one set of studies,  
it was shown that, after a very short time span (a few Monte Carlo (MC) steps for the simulated cell), a paramagnetic (PM) 
configuration quenched sub-critically looks like critical percolation. More precisely, the morphological and statistical 
properties of structures (areas of domains, lengths of interfaces, etc.) that are larger than the typical ones  
($\xi^d(t)$, $\xi^{d-1}(t)$, etc.) are the ones of site percolation at its threshold~\cite{Arenzon07,Sicilia07}. In the other set of studies,
simulations of zero-temperature quenches demonstrated that the systems 
often block into stripe states~\cite{BKR,Olejarz,BP}.
The probabilities of reaching such states were shown to 
equal, with high numerical precision, the ones of having a spanning cluster at critical percolation~\cite{BKR,Olejarz}.
As  the occupation probability for up
and down spins in a high-$T$ equilibrium configuration is smaller than the one at critical percolation,
this fact suggests that the system must have reached
critical percolation at some point. 

Initial states linked to a different equilibrium fixed point are the critical temperature ones.  Studies of quenches from 
$T_c$ into the ordered phase along the lines above were also performed. 
Large structures are, in this case, characteristic of the critical Ising point~\cite{Arenzon07,Sicilia07}. Moreover, the blocked states
reached at $T=0$ appear with probabilities dictated by the ones of the critical Ising spanning structures~\cite{BP}. 

In this work we analyze the transient regime between the initial condition and the  state that will actually control the 
large scale properties in the scaling regime
and the eventually frozen asymptotic configurations. We study the $2d$IM on different lattices. High-$T$ and
critical Ising initial conditions are considered. We determine  the dependence of the time-scale needed to reach this state,
that we call $t_p$, on the lattice coordination and the system size by analyzing how $t_p$ manifests itself in
a number of non-trivial observables. We conjecture its dependence
on the microscopic dynamics. Importantly enough, we generalize the dynamic scaling framework to 
include the influence of  this time-scale, so-far ignored, on the dynamic correlations. 
Finally, we discuss the possible implications of our results on 
systems with more complex out of equilibrium dynamics.

We consider the ferromagnetic Ising model defined by
\begin{equation}
H=-J \sum_{\langle ij\rangle} S_i S_j,
\label{H}
\end{equation}
where the sum is restricted to nearest neighbors on a $2d$ lattice.
We fix $J=1$, and $S=\pm 1$.
The system undergoes a continuous phase transition at a lattice-dependent critical temperature $T_c$.
We consider square, triangular, kagome and bowtie-a lattices with linear size $L$ and either  
free (FBC) or periodic (PBC) boundary conditions.
%~\cite{tilings}.
The coordination numbers are  $n_c^\square = 4$, $n_c^\triangle=6$, $n_c^K= 4$, and $n_c^{\bowtie} = 5$ and the site percolation 
thresholds are  $p^\square_c \simeq 0.59$, $p^\triangle_c = 0.5$, $p^K_c \simeq
0.65$ and $p_c^{\bowtie} \simeq 0.55$. $n_c^{\bowtie}$ is a mean coordination number
since the sites have either $4$ or $6$ neighbors on this lattice.
%~\cite{exact-perc-prob}.
(Regularly odd-coordinated lattices are not suited for our study because the evolution rapidly freezes due
to metastable droplets~\cite{droplets}; lattices such that $p_c<1/2$ are not either because the initial state is super-critical.) 

\begin{figure}[t]
\begin{center}
\includegraphics[scale=0.275]{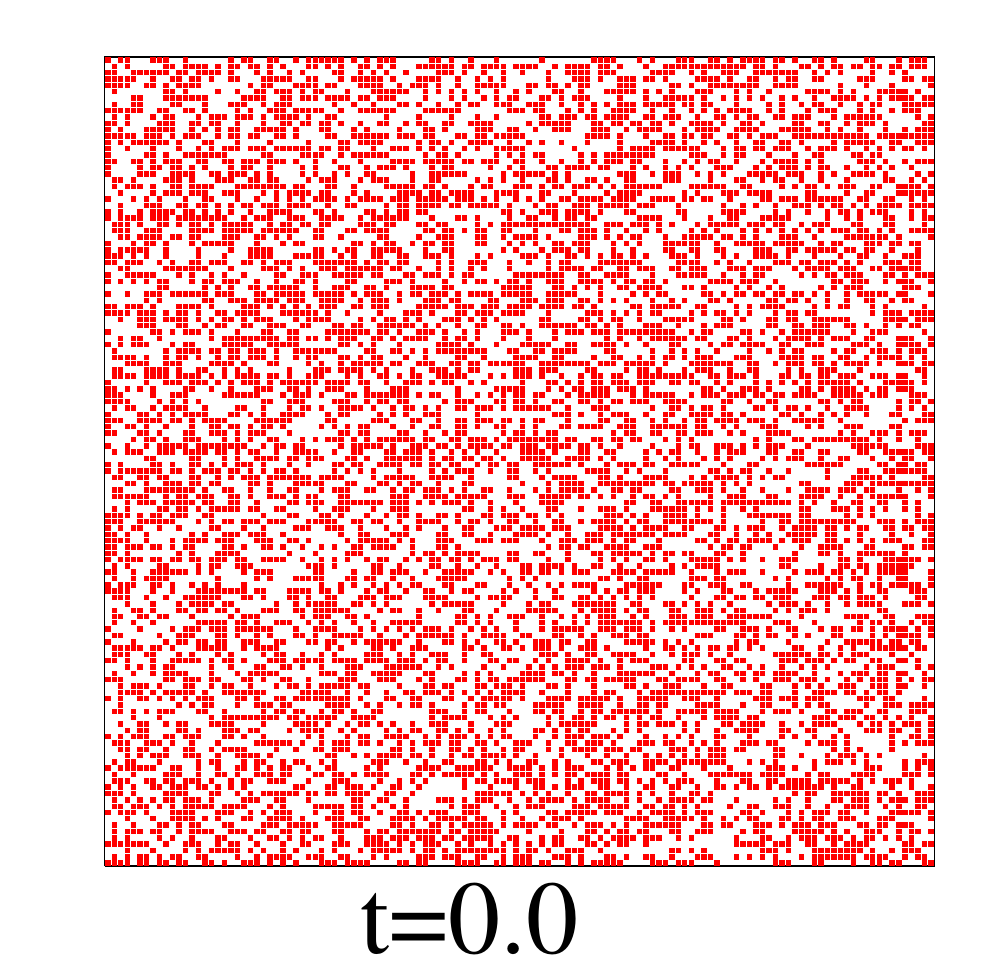}
\includegraphics[scale=0.275]{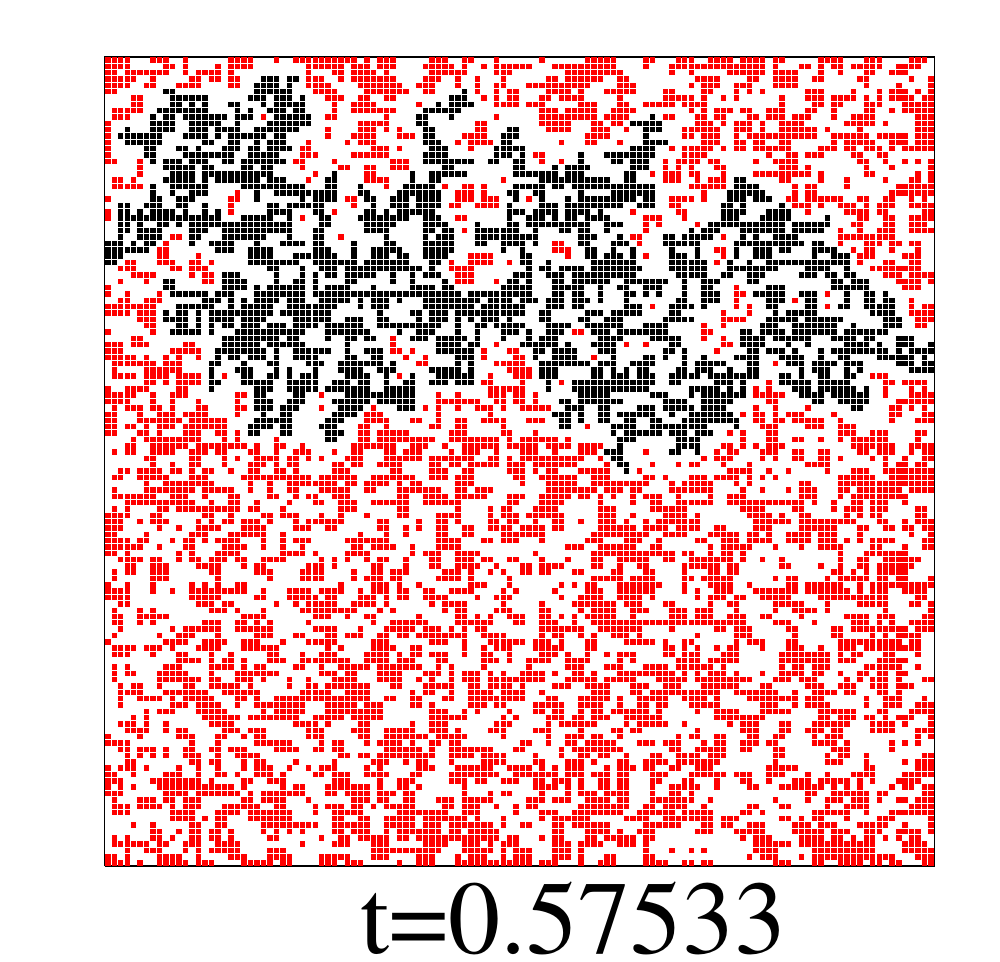}
\includegraphics[scale=0.275]{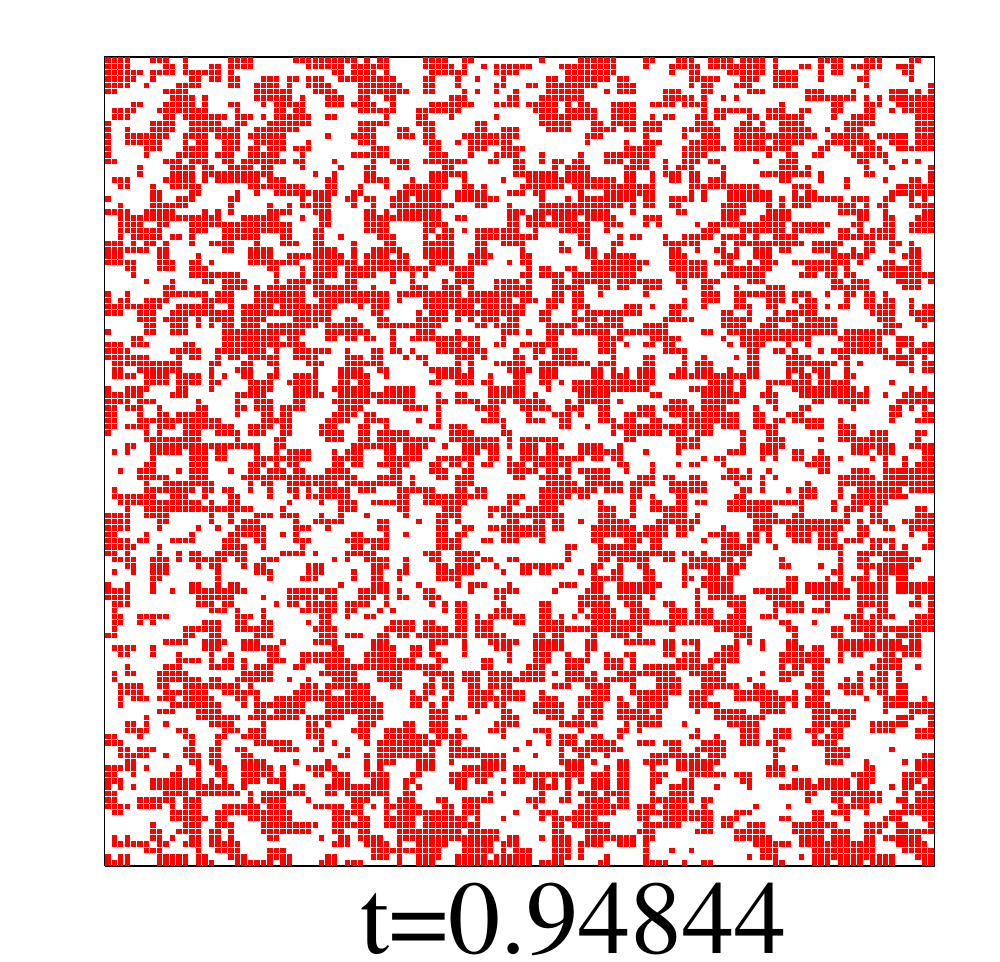}
\includegraphics[scale=0.275]{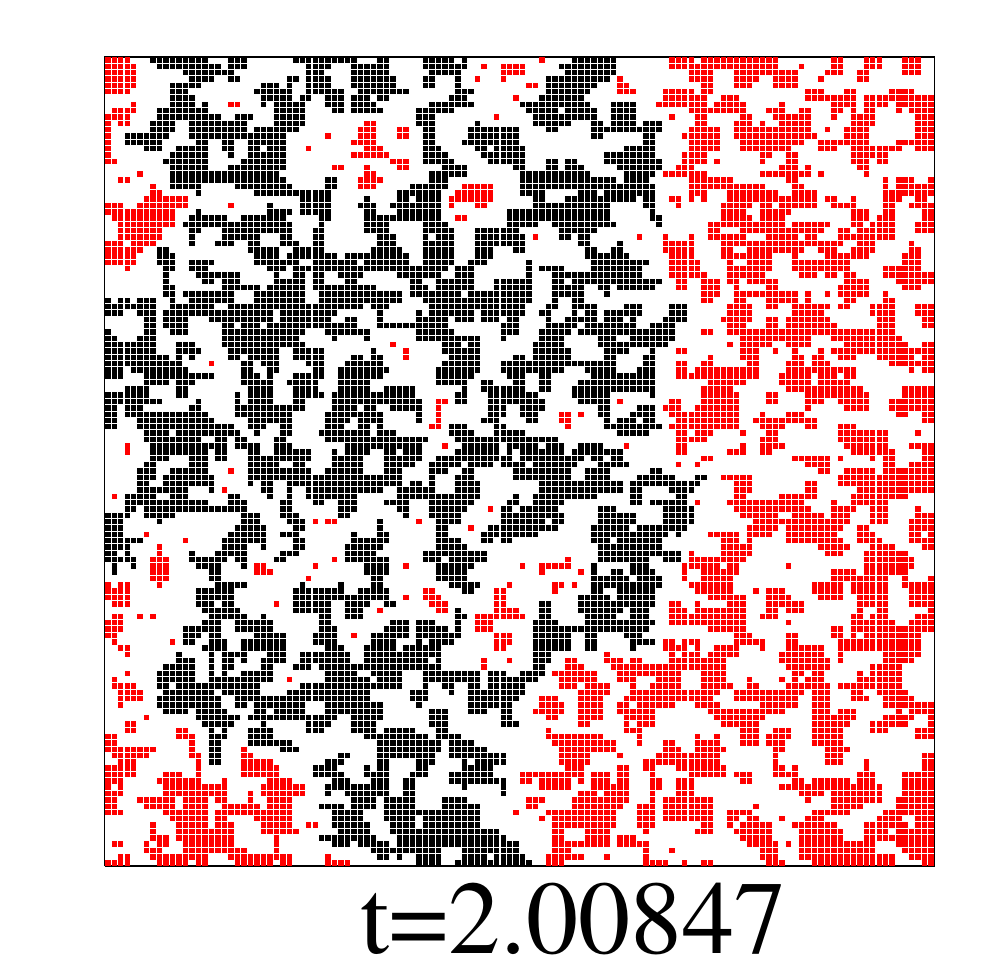}
\includegraphics[scale=0.275]{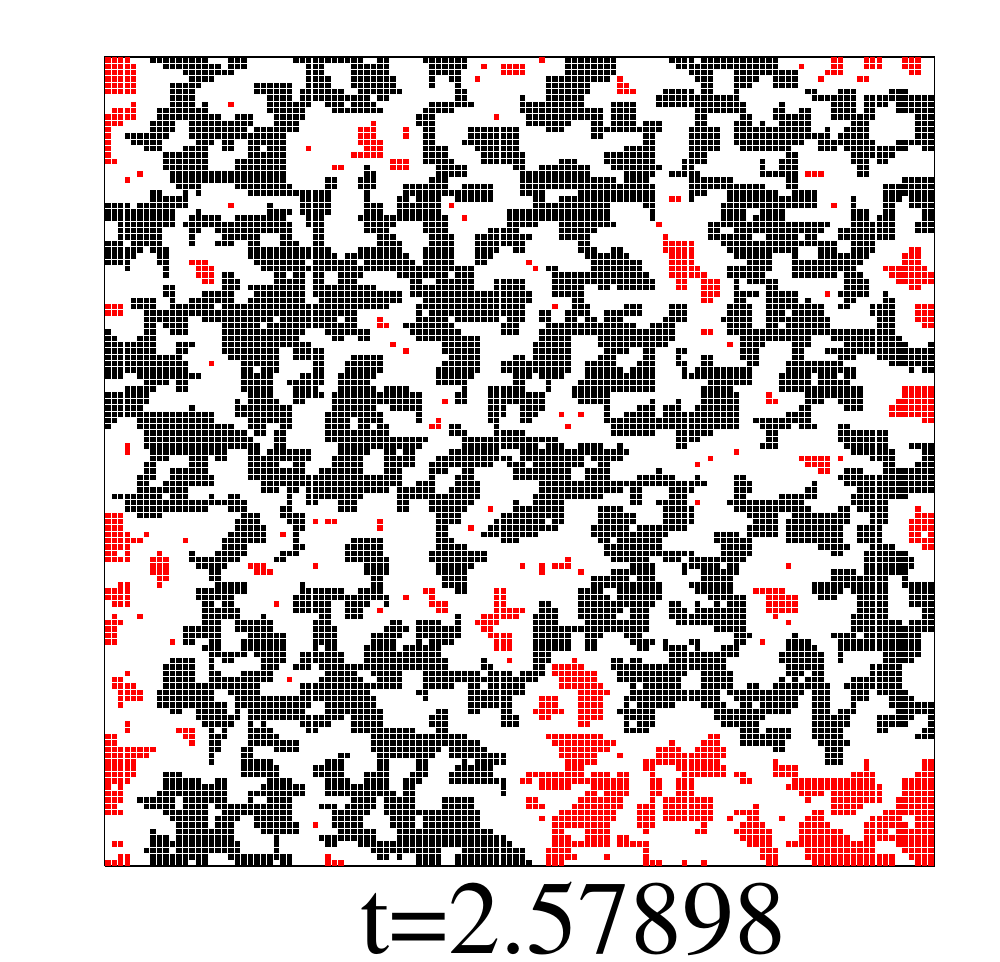}
\includegraphics[scale=0.275]{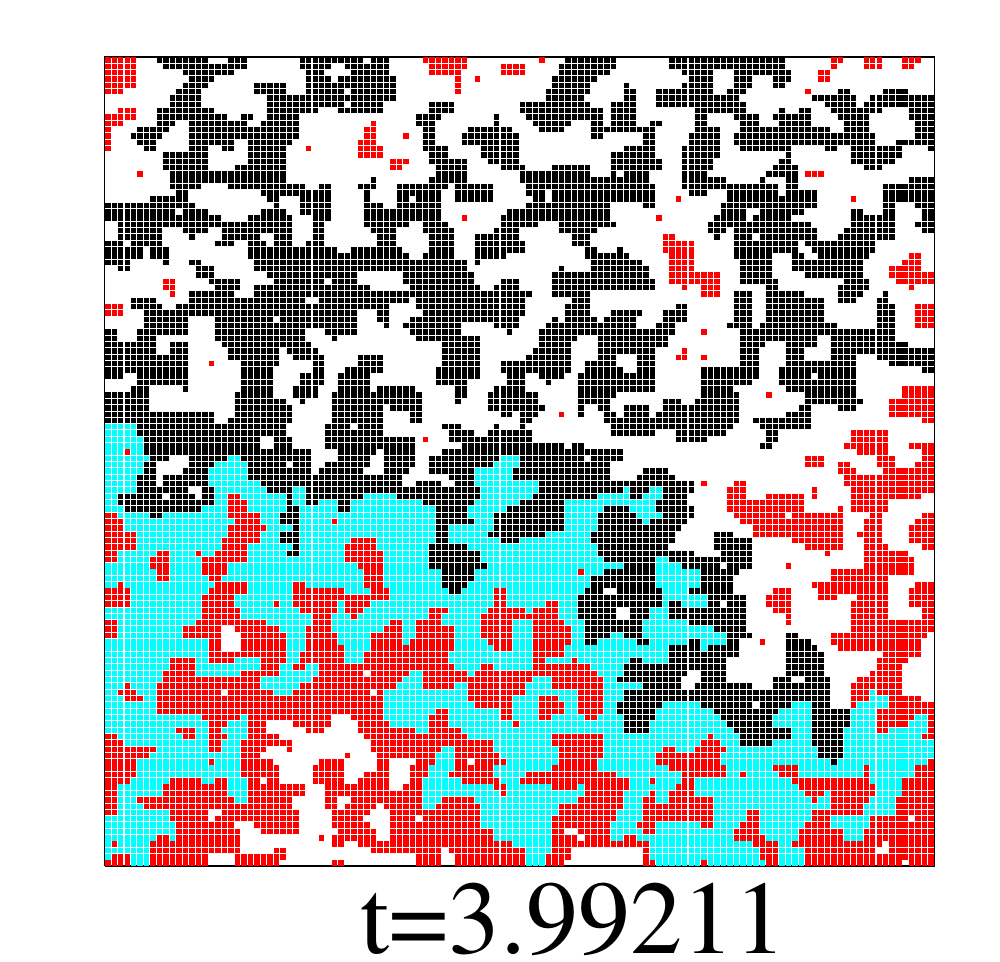}
\includegraphics[scale=0.275]{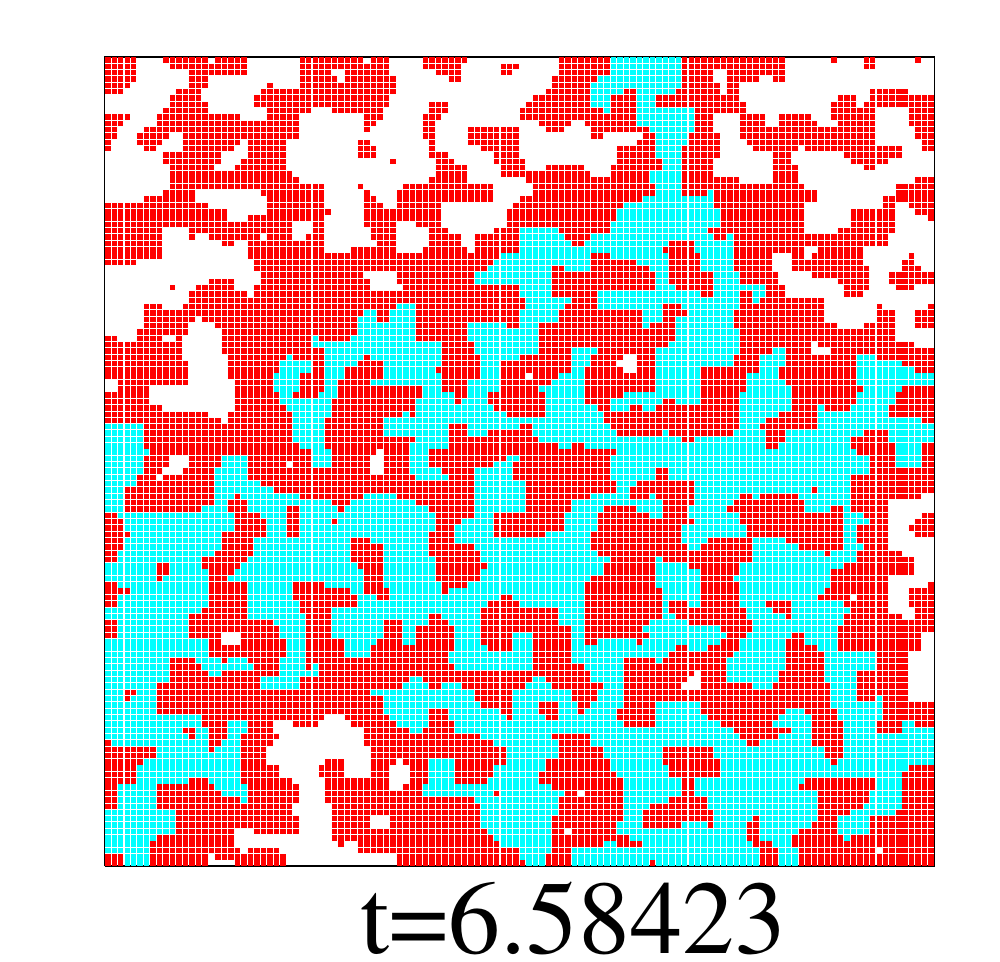}
\includegraphics[scale=0.275]{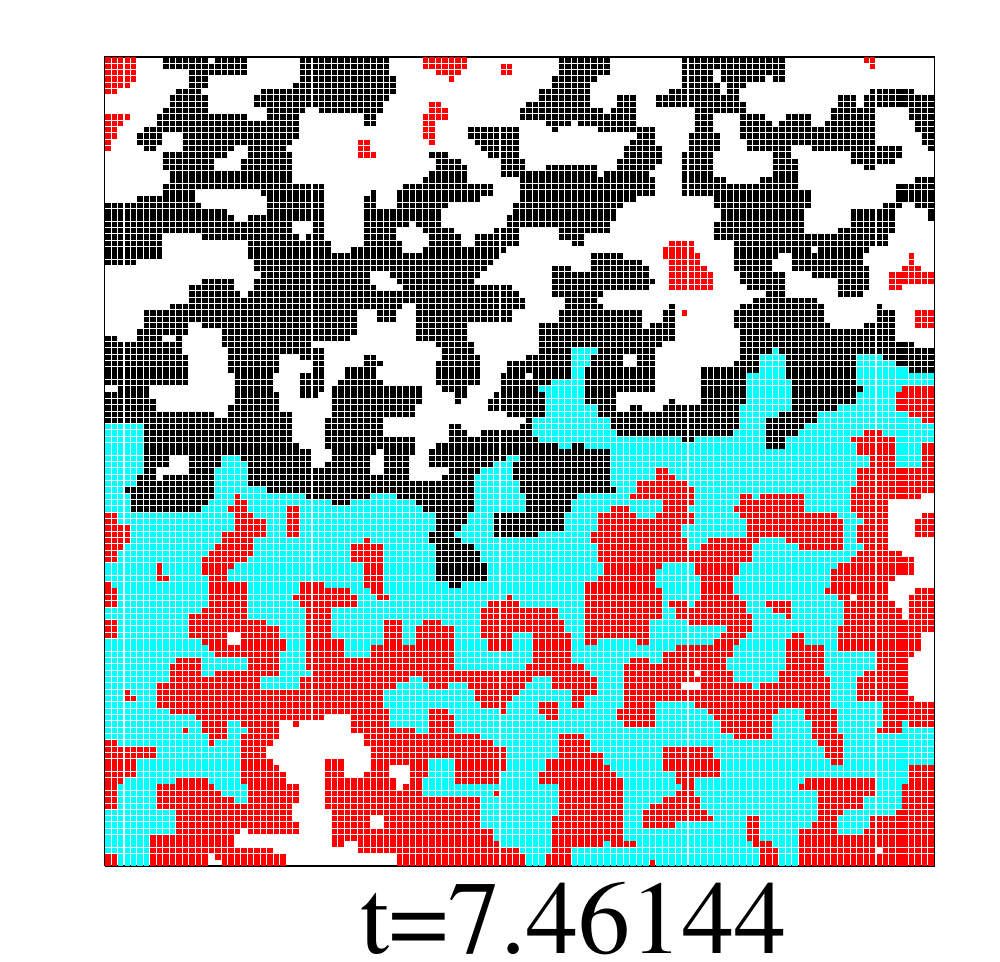}
\includegraphics[scale=0.275]{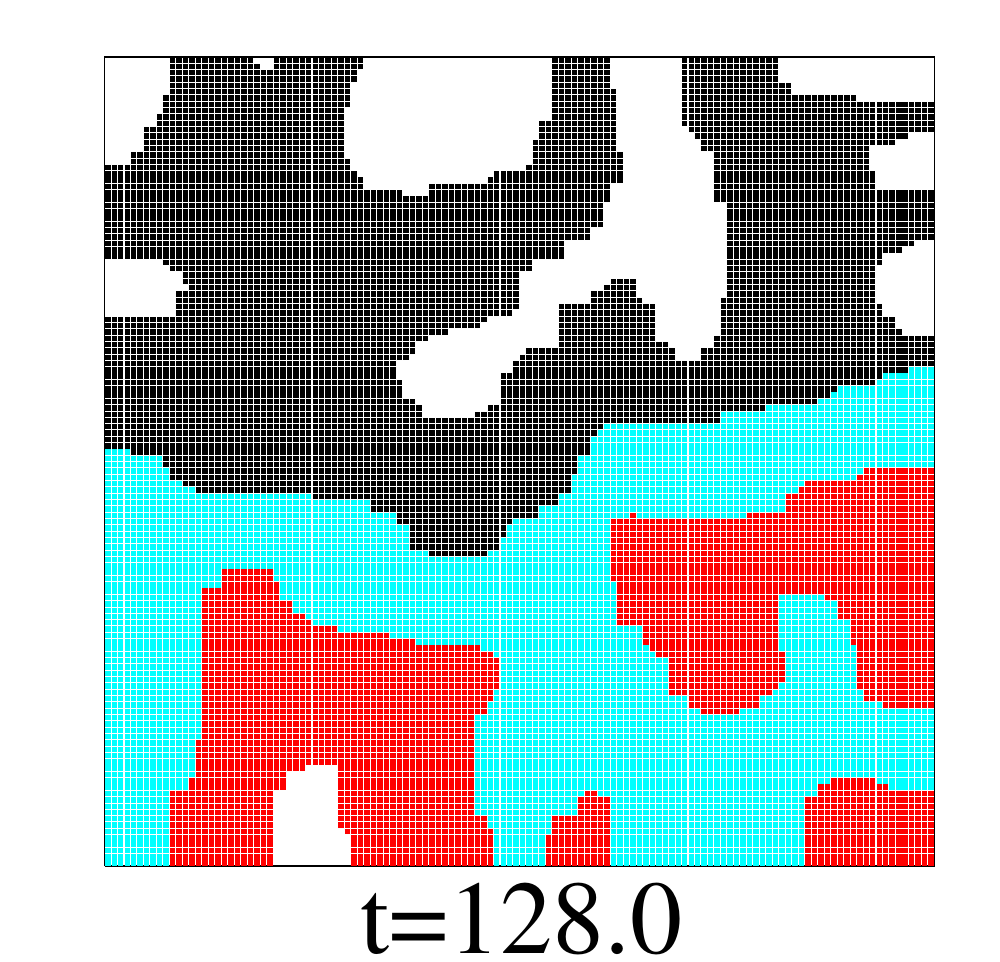}
%\vspace{-2em}
\end{center}
\caption{(Color online.)
Snapshots of a $2d$IM on a square lattice with $L=128$ 
and FBC. A quench from $T_i \to\infty$ to $T=0$ was performed at $t=0$. 
Spins $S_i=-1 \ (S_i= +1)$ are shown in red (white). 
A percolating cluster of spins $S_i=-1$ ($S_i= +1$) is shown in black (clear blue).
}
\label{Percolation}
\end{figure}

We focus on instantaneous quenches from infinite to a sub-critical temperature. The initial PM 
state is a random configuration with $S_i = \pm 1 $ with probability a half.  
%The dynamics are particularly simple at $T=0$. After choosing one site at random, the spin on this site
%is oriented along the local field (which is the sum of the nearest neighbor spins). If this local field is
%zero, the value of the spin is chosen randomly.  
The dynamics follow the MC rule. $L^2$ spin-flip attempts correspond to $\delta t =
1$. At $T<T_c$, the number of flippable spins decreases with time
% and testing all spins in the sample results in a waste of computer time. 
so we accelerate our simulations with a Continuous Time MC
algorithm~\cite{Bortz}. 
Unless otherwise stated the number of samples used is at least $10^6$.
%that considers only the spins that can be reversed (at generic $T$). 

The equilibration time $t_{eq}$ is estimated from $\xi(t_{eq})\sim L$ and yields $t_{eq}\sim L^2$. After $t_{eq}$, 
the configuration is in one ergodic component, or it is
%, i.e. all the spins take the same value, or it is
a stripe state with interfaces crossing the lattice. The stripes are stable or not depending on the lattice and 
boundary conditions. Unstable
stripes are destroyed via a different mechanism over a much longer time-scale~\cite{lipowski,SKR1SKR2}.

Let us start by analysing the dynamics of the square lattice model.
A glimpse on a set of snapshots taken at and after the quench shows that, although there are no 
percolating clusters in the initial state, these appear very soon. Figure~\ref{Percolation} displays the configurations of a 
system with $L=128$ and FBC quenched from $T_i\to\infty$ to $T=0$. 
At $t=0$ the spins take random values and no spanning cluster is present. At  $t_f\simeq 0.57\ll
L^2$ a first spanning cluster appears. Then follows a period  $t_f<t<t_p$  with $t_p\simeq 7.46\ll L^2$ in which spanning
clusters appear and disappear typically $10$ to $20$ times in our simulations.  At $t_p$ the number and type of spanning clusters 
are equal to the number and kind of stripes in the final state. 
The fact that $t_p\ll L^2$ is intriguing as it indicates that the fate of the Ising model is sealed very
soon in the evolution.
The regime $t_f<t<t_p$ is also interesting but  we will not consider it here.

\begin{figure}[t]
\begin{center}
\includegraphics[scale=1.1]{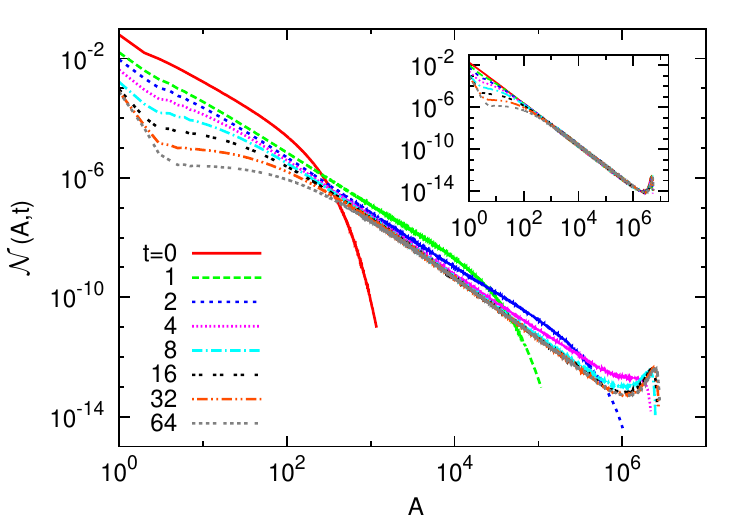}
\vspace{-1em}
\end{center}
\caption{${\cal N}(A,t)$ vs. $A$ at different times $t$ given in the key after a quench from $T_i\to\infty$ to $T_c/2$ for
	a system with  $L=2560$. In the inset  data obtained after a quench from $T_c$ to $T_c/2$.
}
\label{NA2560}
\end{figure}

A first quantitative indication of critical percolation playing a role is given by the probability
distribution of spin cluster areas, ${\cal N}(A,t)$, as a function of the number of spins in a cluster, $A$,
and time, $t$~\cite{Arenzon07,Sicilia07}.
At $t=0$ the system is not critical and ${\cal N}(A,0)$ decays exponentially with $A$, as shown by the solid (red) curve
in the main panel in Fig.~\ref{NA2560}.  After a very short time, $t \simeq 10$, the tail tends to an algebraic decay
${\cal N}(A,t) \simeq 2c A^{-\tau}$. For finite $L$ and sufficiently long times the power-law is cut-off by a bump 
around areas that scale with $L$. For $t \gtrsim 16$ 
the power-law holds in the range $10^3 \leq A \leq 10^6$ and the bump is around
$A \simeq 2 \ 10^6$.
The functional form
\begin{equation}
\label{dist}
{\cal N}(A,t) \simeq 2cA^{-\tau} +  N_p(A/L^{D},t)
\end{equation}
takes into account all these features.  $D=d/(\tau-1)$ is the fractal dimension of the percolating clusters.
%We also note that for the areas subject to the curvature driven dynamics, $A \lesssim \xi^2(t)$ 
%(that do not interest us here), the curves decrease with $t$~\cite{Arenzon07,Sicilia07}. 
%=d/(\tau-1)=91/48\simeq 1.896$ their fractal dimension.

A direct fit of the power-law
yields $\tau = 2.020 - 2.040$ depending on the fitting range.  While these values are close to the
percolation one, $\tau =187/91 \simeq 2.05495$, they are also close to the critical $2d$IM one,
$\tau = 2.02674$, and it is difficult to distinguish between these two cases with this measurement
(having said this, the numerical value of $c$ is very close to the analytic one for percolation, 
$c=1/(8\sqrt{3}\pi)$~\cite{Cardy} and distinctively different from the one for the critical IM~\cite{Arenzon07,Sicilia07}). 

\begin{figure}[t]
\begin{center}
\includegraphics[scale=0.675]{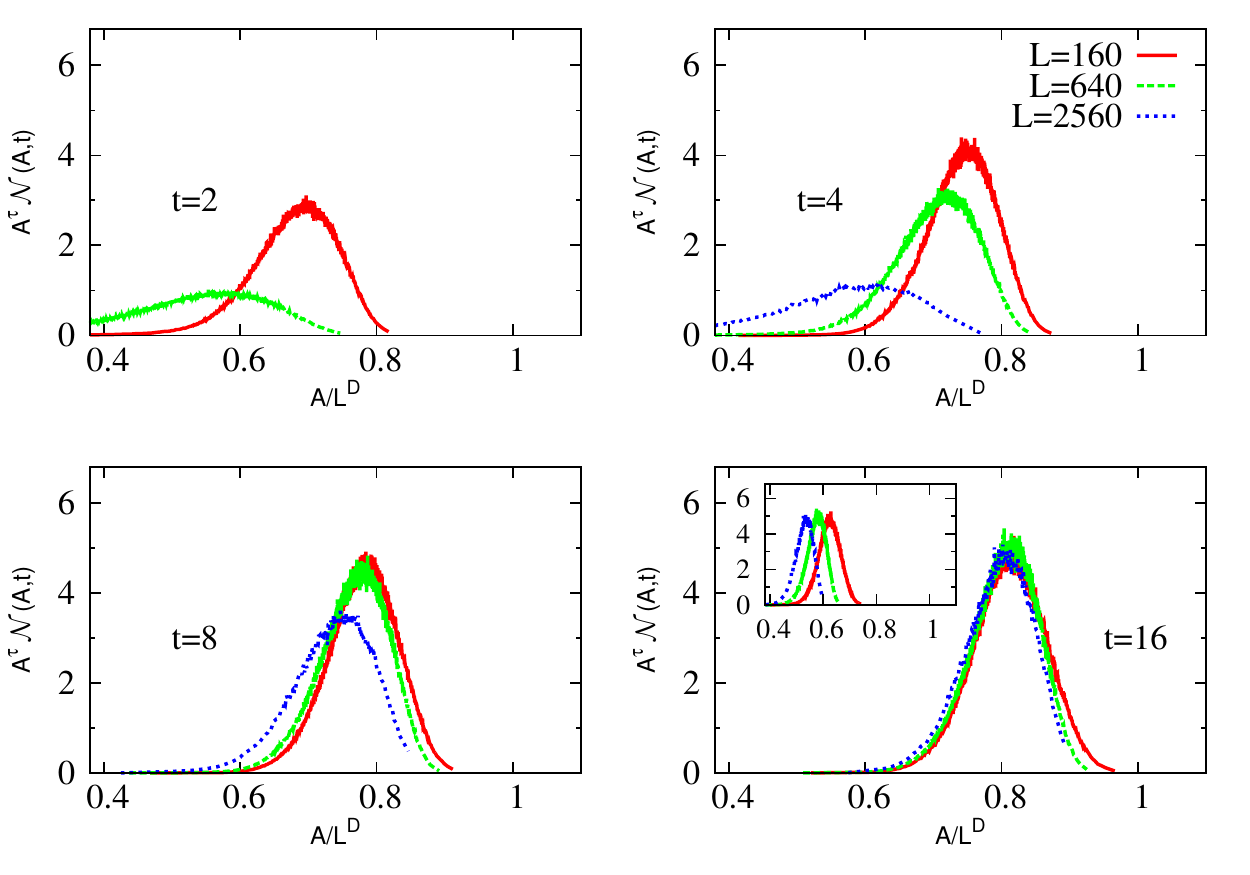}
\vspace{-1em}
\end{center}
\caption{(Color online.)
$A^{\tau} {\cal N}(A,t)$ vs. $A/L^{D}$ at different times $t$ after a quench from $T_i\to\infty$ to 
$T_c/2$. We employ the parameters $\tau$ and $D$ of critical percolation 
except in the inset of the lower right panel where we use the ones of the critical Ising model. 
The number of samples used to build the pdf is $10^6$ for $L=160$ and $L=640$ and $2 \ 10^5$ for $L=2560$.}
\label{QuenchTI} 
\end{figure}  

The  study of the bump, scaled as in the second term in~(\ref{dist}), lifts all ambiguities on the 
value of $\tau$. In Fig.~\ref{QuenchTI}
we plot  $A^{\tau} {\cal N}(A,t)$ vs.
$A/L^{D}$ with the percolation $\tau$ for  $t \in [2,16]$ and different  $L$.  For small $t$ 
the height of the distribution depends on $L$ while for $t \gtrsim 16$ they all collapse. 
 The inset in the lower right panel 
displays the same data scaled with the critical IM value of $\tau$.  Clearly, there is no scaling with this choice. 

We note that the time required for the distribution to become percolation-like increases  weakly with $L$. 
The curves for $L=160$ and $L=640$ at $t=2$ are replaced by the curves for
$L=640$ and $L=2560$ at $t=4$. The same holds true between $t=4$ and $t=8$. At $t=8$, the curves for $L=160$ and $L=640$
do collapse, as for data at $t=16$ for $L=640$ and $L=2560$. This suggests that there is a time scale 
%\begin{equation}
$t_p \sim L^{\alpha_p}$, 
%\end{equation}
 with $\alpha_p \sim 0.5$, after which the bump remains stable in the scaling plot.
It is still not clear whether this bump is made of everlasting percolating structures 
or whether these still appear and disappear as illustrated in Fig.~\ref{Percolation}.

The time-scale $t_p$ after which the actual percolating structure stabilizes
can be estimated from the analysis of the correlation of the number of crossing
clusters present at time $t$ and those surviving in the blocked configuration 
at zero temperature:
\begin{equation}
{\cal A}_c(t) = \langle \delta_{n_v(t)n_v^\infty}\delta_{n_h(t)n_h^\infty}\rangle \; .
\label{eq:crossing-corr}
\end{equation}
We note $n_h(t)$ [$n_v(t)$] the number of horizontal
(vertical) crossing clusters at time $t$, $n_h^\infty \ (n_v^\infty)$ the number of 
horizontal (vertical) spanning clusters  in the blocked configuration,
%The total number of crossing clusters $n_c(t)=n_h(t)+n_v(t)$.  
$\langle \cdots \rangle$ the average over initial configurations and thermal histories, and $\delta_{ab}$ 
the Kronecker delta. 
For geometrical reasons the clusters crossing in only one direction must be at least two in the final state so that if $n_h(t)=0$ and $n_v(t)>0$ then
$n_v(t)\geq 2$. A cluster crossing in both directions is always unique and $n_h(t)=n_v(t)=1$. For a
configuration with no crossing clusters  $n_h(t)=n_v(t)=0$.  

\begin{figure}
\begin{center}
	\includegraphics[scale=0.9]{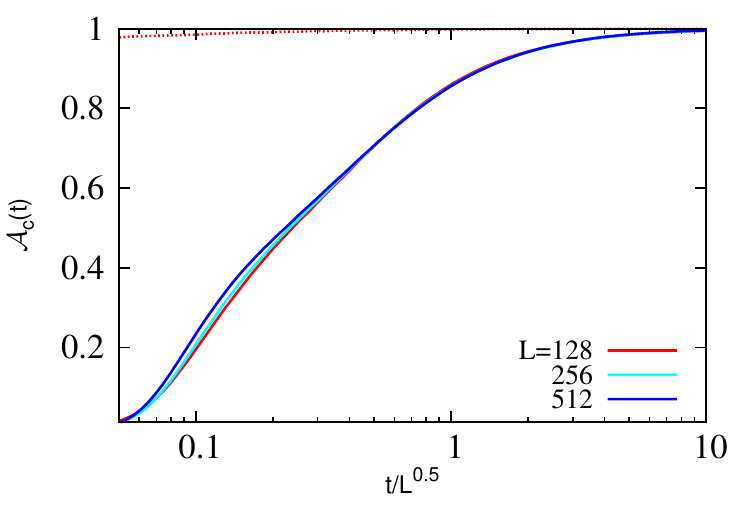}
\vspace{-1.5em}
\end{center}
\caption{(Color online.) Number of 
crossings autocorrelation, ${\cal A}_c(t)$ defined in Eq.~(\ref{eq:crossing-corr}), after a quench of the  FBC square lattice
form $T_i\to\infty$ to $T=0$.
Data are displayed as a function of $t/L^{0.5}$ in linear-log scale for the values of $L$ given in the key. 
The dashed red line very close to $1$ is the behavior of ${\cal A}_c(t)$ for a quench from $T_c$ to $T=0$ 
for $L=128$.} 
\label{Overlap3}
\end{figure}

The correlation ${\cal A}_c(t)$ interpolates between $0$ and $1$. For sufficiently large system sizes, $n_h(0)=n_v(0)=0$.
In the blocked configuration there should be at least one crossing cluster, leading to $n_h^\infty\neq 0$ and/or $n_v^\infty\neq 0$.  
In Fig.~\ref{Overlap3} we show ${\cal A}_c(t)$ on a square lattice with 
FBC. 
%We use a log-linear scale with the scaling variable $t/L^{0.5}$ in the horizontal-axis.  
A very accurate data collapse 
for different sizes is found for $t/L^{0.5} \gtrsim 0.1$.   At shorter times  the scaling is not as good 
due to the large number of states with only one cluster percolating in one direction
(see Fig.~\ref{Percolation}) present on the square
lattice because of interface corners. As soon as interfaces become flat these configurations no longer exist.  The convergence to
$1$ is fast since ${\cal A}_c(t)\sim 1$ for  $t/L^{0.5}>10$. This indicates that $t_p$
scales as $t_p\sim L^{0.5}$ on the square lattice.

\begin{figure}[ht]
\begin{center}
\includegraphics[scale=1.05]{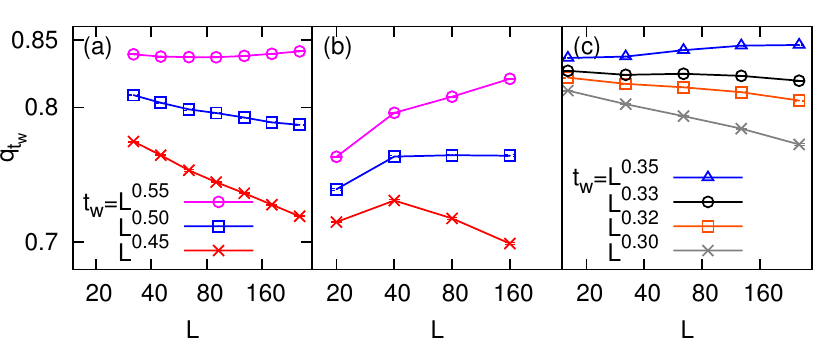}
\vspace{-2em}
\end{center}
\caption{
(Color online.) The  $L$-dependence of the asymptotic overlap $\lim_{t\to\infty} q_{t_w}(t,L)$ 
in the $2d$IM quenched  from a PM state to $T=0$ at $t=0$.
(a) Square lattice with FBC.
(b) kagome lattice with FBC.
(c) Triangular lattice with PBC.
The values of $t_w(L)$ are given in the keys. Data have been averaged over $10^6-10^7$ samples.
}
\label{Overlap2}
\end{figure}

The overlap between two copies of the system~\cite{Derrida87,Cude,Barratetal,Machta13} defined as follows also informs us about the time $t_p$. 
At $t=t_w$ one makes two copies of the configuration, say $s_i(t_w)=\sigma_i(t_w)$, 
and lets them evolve with different noises. The overlap between the  clones at   time $t$ is
\begin{equation}
q_{t_w}(t,L) = \frac{1}{N} \sum_i \langle s_i(t) \sigma_i(t)\rangle
\; , 
\end{equation}
where the angular brackets indicate an average over different realizations of this 
procedure. The two-time scaling properties of this quantity, in the limit $L\to\infty$, 
were used to distinguish domain-growth processes from glassy ones~\cite{Cude,Barratetal}. 
 It was recently shown that $q_{t_w=0}(t,L)$ and $q_{t_w=0}(\infty,L)$ decrease algebraically with $t$ and 
 $L$, respectively~\cite{Machta13}. 
The fact that $q_{t_w=0}(t,L)$ decreases with $t$ is clear.  Even though we know that there are 
a finite amount of stripes in the final state~\cite{BKR,Olejarz}, these are not encoded in the initial condition
and there is no reason why the different thermal histories will take the two clones to  the same
percolating state.   Instead, by letting $t_w$ go beyond $t_p$, the two clones should be 
strongly correlated for all subsequent times since they are in the same
percolating state. We set this argument in practice by computing $\lim_{t\to\infty} q_{t_w}(t,L)$ for various $t_w(L)$. 
The outcome is shown in  
Fig.~\ref{Overlap2} for the square (a), kagome (b), and triangular $($c$)$ lattices. For the square and kagome lattices, we observe that if 
$t_w(L)$ increases as $L^{0.5}$, the overlap remains 
constant with a finite value, thus confirming that $t_p \sim L^{0.5}$ in these cases.

On the triangular lattice $p^\triangle_c=1/2$ and there is a percolating cluster in the initial 
condition with non-zero probability. A naive guess would be that this percolating state survives after the
quench, and  $t_p\equiv 0$. This, however, is not true as $q_0(t,L)$ decays to zero in the large size and time 
limit~\cite{bpc}. As shown in panel (c) in Fig.~\ref{Overlap2}, only for $t_w > t_p \simeq L^{0.33}$, $\lim_{t\to\infty} q_{t_w}(t,L)$ saturates.
The initial state, although percolating, is not stable under the 
dynamics and a transient scaling with $L$ is still needed to reach the truly stable one~\cite{supplement}.
We have also simulated the bowtie-a lattice (not shown) and found
$t_p\sim L^{0.38(5)}$. We have checked that in all the lattices studied the constant prefactor in $t_p$ is of order $1$.

These results suggest that, after a sub-critical quench from $T_i\to\infty$,  
\begin{equation}
t_p \simeq L^{\alpha_p} 
\; , 
\;\;\;\;
\;\;\;\;
\alpha_p = z/n_c
\; , 
\label{eq:prediction}
\end{equation}
with $z$ the dynamic exponent and $n_c$ the regular or averaged lattice coordination. 
%The fact that $t_p$ diverges 
%with $L$ means that for infinite size the systems
%never reach a definitive percolating state. However, 

For the system sizes commonly used in the literature $L\simeq 100-1000$,
 $t_p \simeq 10-30$, and percolation is very quickly attained.   It is important to notice 
 that, since the equilibration time remains at least  
$t_{eq}\sim L^2 \gg t_p$~\cite{Bray94}, or even longer due to blocked states~\cite{lipowski,SKR1SKR2}, 
after stable percolation of an ordered cluster is established 
at $t_p$ the systems are still far from equilibrium, as shown by 
 the correlation and response functions that continue to relax 
well beyond this time-scale~\cite{Barrat,Corberi}.
The new feature provided by our study is that the scaling properties are modified by 
the {\it extra} time-scale $t_p$.
Indeed, according to the usual formulation of the dynamical scaling hypothesis,
in the regime $\xi _{micro}\ll \xi (t) \ll L$, when $\xi$ is grown much larger than
a microscopic length $\xi _{micro}$ associated to the lattice spacing but is still
much smaller than the system size when equilibration or blocked states start to occur,
the statistical properties  do not depend on time provided that distances
are measured in units of the dominant length $\xi (t)$. Due to this fact,
correlators such as $G(r,t)=\langle S_i(t)S_j(t)\rangle$, where $r=|i-j|$, 
take the scaling form
\begin{equation}
G(r,t,L)=f\left [\frac{r}{\xi(t)}\right ],
\label{scalG}
\end{equation}
where $f(x)$ is a scaling function,
expressing the fact that there is a unique relevant length in the system.
However, the presence of $t_p$ introduces another characteristic length 
${\cal L}(L)=\xi(t_p)\simeq L^{\alpha _p/z}$
separating an early stage in which there are no stable percolating structures
from a late stage in which they exist. In the presence of two characteristic 
lengths, since ${\cal L}$ plays a companion role to that of $\xi$, 
the proper scaling form for $G(r,t,L)$ should read 
\begin{equation}
G(r,t,L)=g\left [\frac{r}{\xi(t)},\frac{{\cal L}(L)}{\xi (t)}\right ]
\label{newscalG}
\end{equation}
with a new, two-variable, scaling function $g(x,y)$. 
The difference between Eqs.~(\ref{scalG}) and
(\ref{newscalG}) is manifest when trying to collapse data for $G(r,t,L)$ at different
times $t$. Indeed, according to Eq.~(\ref{newscalG}) curves 
for a system of a given size $L$ (and hence a given ${\cal L}$)
cannot be superimposed on a master-curve by simply plotting them against $r/\xi (t)$,
as Eq.~(\ref{scalG}) would suggest, because in so doing  the second argument
entering the scaling function $g$ changes. Instead, Eq.~(\ref{newscalG}) states
that it is possible to collapse curves
at different times $t=t_0,t_1,\dots,t_i,\dots$ if they are relative to systems of 
different sizes $L=L_0,L_1,\dots,L_i,\dots$ chosen in such a way that 
${\cal L}(L_i)/\xi(t_i)=const.$, by
still plotting them against $r/\xi(t)$. 

\begin{figure}[t]
\begin{center}
\includegraphics[scale=1.0]{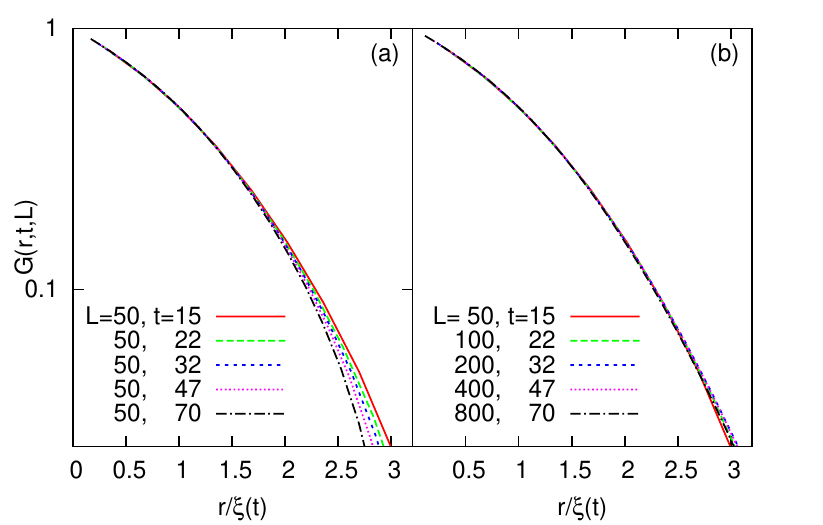}
\end{center}
\vspace{-0.25cm}
\caption{
(Color online.) $G(r,t,L)$ against $r/\xi(t)$  
in the $2d$IM quenched  from a PM state to $T=0$ at $t=0$.
The system size and measuring times are given in the keys.
Data have been averaged over $2\cdot 10^4$ samples.
}
\label{fig_scal}
\end{figure}

In order to check this we have computed 
$G(r,t,L)$ on square lattices of sizes $L_i=2^i \cdot L_0$, with $L_0=50$ and 
$i=0-4$. Using Eq.~(\ref{eq:prediction}) and $\xi(t)\sim t^{1/z}$ leads to
${\cal L}(L_i) \simeq L_i^{1/n_c}$. We have considered 
times $t_i$ such that $t_0=15$ and ${\cal L}(L_i)/\xi(t_i)\simeq 16.86$.
Having verified that, for these times, the scaling condition   
$\xi _{micro}\ll \xi (t) \ll L$ is met for any $L_i$,
we try to collapse the data for 
the smaller system according to Eq. (\ref{scalG}), namely by simply plotting the data
against $r/\xi (t)$ (where $\xi$ is obtained as the half-height width of 
$G$, namely from the condition $G(\xi ,t)\equiv 1/2$). 
The result is shown in Fig.~\ref{fig_scal}~(a), where
one observes a systematic
downward spreading of the curves from $r/\xi (t)\simeq 1.5$ onwards (the collapse 
 of the curves at $r/\xi=1$ is exact due to the operative definition of
$\xi(t)$). In Fig.~\ref{fig_scal}~(b), instead, we plot the data
at the same times $t_i$ but for  systems with different sizes $L_i$.
As expected, the quality of the collapse is, in this case, much better.
Besides shading some light on the scaling properties of coarsening systems,
this results confirm  the validity of Eq. (\ref{eq:prediction}) in an independent way.

We repeated the analysis for various choices of the constant ${\cal L}(L_i)/\xi(t_i)$ finding that, in all cases,
Eq.~(\ref{newscalG}) provides a better description of data than Eq.~(\ref{scalG}), although
it is clear that for times $t\gg t_p$ (so that $\xi(t)\gg {\cal L}(L)$) the validity
of Eq.~(\ref{scalG}) gets progressively restored since one expects 
$\lim _{y\to 0} g(x,y)\simeq g(x,0)=f(x)$. 
We have also checked that, similarly to what happens for $G$,
a better scaling description of other 
correlators, such as, for instance, the autocorrelation function 
$C(t,t_w,L)=\langle S_i(t)S_i(t_w)\rangle$, can be obtained by taking into account the
presence of $t_p$. Indeed, also for this quantity we have found that
a two-variable scaling form 
$C(t,t_w,L)=h\left [ \frac{\xi (t_w)}{\xi (t)},\frac{{\cal L}(L)}{\xi(t)}\right ]$
improves the collapse with respect to the usually conjectured form where the role
of the second argument in $h$ is neglected.

The situation is different if we start from a configuration equilibrated at $T_c$.  
The inset in Fig.~\ref{NA2560} shows
${\cal N}(A,t)$ after a quench from $T_c$ to $T_c/2$. In this case, the initial state is critical and the curve at $t=0$ is 
a power law~\cite{Arenzon07,Sicilia07}.  Again, extracting $\tau$ from it is hard. Instead,  the bump described by the second term in (\ref{dist}) 
can only be scaled by using the  critical IM $\tau$ (not shown). Moreover, the crossing properties of the final state are already 
contained in the initial configuration~\cite{BP}. ${\cal A}_c(t)$ is immediately close to one, see the red
dashed line in Fig.~\ref{Overlap3},
as well as the clone overlap $q_{t_w}(t,L)$ (not shown).

These results raise many questions. We expect the effect of the working temperature to be weak and 
alter only the pre-factor in (\ref{eq:prediction}). The effect that the microscopic dynamics (conserved 
quantities) may have on $t_p$ remains to be clarified. Weak quenched disorder slows down the
curvature driven growth. Does it also affect the transient regime? 
%Our results for three lattice geometries with uniform coordination are compatible with (\ref{eq:prediction}). 
%It is however difficult to put this guess to the test on
%other such lattices since they are not so numerous. Instead, one could repeat this analysis 
%on non-uniform (non-Archimedean) lattices and test whether $\alpha_p=0.5$ for all of them apart from
%the special triangular case in which there is a spanning cluster from the start.
It would also be interesting to pinpoint the implications of these results in higher dimensions. 

It should be possible to search for percolation effects experimentally in systems undergoing 
phase ordering kinetics for which visualization techniques
have proven to be successful. Two examples are liquid crystals~\cite{Xtals} and phase separating glasses~\cite{Damien}.


\begin{thebibliography}{99}

\bibitem{Bray94} A. J. Bray,
Adv. Phys. {\bf 43}, 357 (1994).

\bibitem{percolation}
D. Stauffer and A. Aharony, 
{\it Introduction to Percolation Theory}, 2nd ed. (Taylor \& Francis, London, 1994).

\bibitem{Arenzon07}
%    Exact results for curvature-driven coarsening in two dimensions
    J. J. Arenzon, A. J. Bray, L. F. Cugliandolo, and A. Sicilia,
Phys. Rev. Lett. {\bf 98}, 145701 (2007),

\bibitem{Sicilia07}
%Domain growth morphology in curvature driven two dimensional coarsening
    A. Sicilia, J. J. Arenzon, A. J. Bray, and L. F. Cugliandolo,
Phys. Rev. E {\bf 76}, 061116 (2007).

\bibitem{BKR}
K.~Barros, P.~L.~Krapivsky, and S.~Redner,
%Freezing into stripe states in two-dimensional ferromagnets and crossing probabilities in critical percolation. 
Phys. Rev. E {\bf 80}, 040101 (2009). 

\bibitem{Olejarz}
J. Olejarz, P. L. Krapivsky, and S. Redner,
Phys. Rev. Lett. {\bf 109}, 195702 (2012).

\bibitem{BP} T.~Blanchard and M.~Picco, 
%Frozen into stripes: fate of the critical Ising model after a quench,
%arXiv:1304.6758.
Phys. Rev. E \textbf{88}, 032131 (2013).

%\bibitem{tilings}
%B. Gr\"unbaum and G. C. Shephard,
%{\it Tilings and Patterns}
%(New York: Freeman \& co., New York, 1987).

%\bibitem{exact-perc-prob}
%M. F.  Sykes and J. W. Essam,
% "Exact critical percolation probabilities for site and bond problems in two dimensions". J
%J. Math. Phys. {\bf 5}, 1117 (1964).
%Journal of Mathematical Physics 5 (8): 1117Ð1127.

\bibitem{droplets}
 %ORDERING PROCESS IN THE KINETIC ISING-MODEL ON THE HONEYCOMB LATTICE
H. Takano and S. Miyashita, 
Phys. Rev. B {\bf 48}, 7221 (1993).

\bibitem{Bortz} A.~B.~Bortz, M.~H.~Kalos, and J.~L.~Lebowitz, 
%A new algorithm for Monte Carlo simulation of Ising spin systems. 
	J. Comp. Phys. {\bf 17}, 10 (1975).

\bibitem{lipowski}
%Anomalous phase-ordering kinetics in the Ising model
A. Lipowski, Physica A,
\textbf{268}, 6--13 (1999).

\bibitem{SKR1SKR2}
V.~Spirin, P.~L.~Krapivsky, and S.~Redner,
Phys. Rev. E {\bf 63}, 036118 (2001),
Phys. Rev. E {\bf 65}, 016119 (2001).
	
\bibitem{Cardy}
J. Cardy and R. M. Ziff, J. Stat. Phys.
{\bf 110}, 1 (2003).

\bibitem{Derrida87}
B. Derrida, 
%Dynamical phase transition in nonsymmetric spin glasses 
J. Phys. A {\bf 20}, L721 (1987).

\bibitem{Machta13}
 %    Nature vs. Nurture: Predictability in Zero-Temperature Ising Dynamics
    J. Ye, J. Machta, C. M. Newman, and D. L. Stein, 
%    arXiv:1305.3667.
Phys. Rev. E \textbf{88}, 040101(R) (2013).

\bibitem{Cude}
L. F. Cugliandolo and D. S. Dean, 
Phys. A {\bf 28}, 4213 (1995).

\bibitem{Barratetal}
A. Barrat, R. Burioni, and M. M\'ezard, 
J. Phys. A {\bf 29}, 1311 (1996).

\bibitem{bpc} T. Blanchard, L. F. Cugliandolo, and M. Picco, in preparation.

\bibitem{supplement} See Supplementary Material for snapshots of the evolution of a system on a triangular lattice.

\bibitem{Barrat}
A. Barrat, Phys. Rev. E {\bf 57}, 3629 (1998).

\bibitem{Corberi}
F. Corberi, E. Lippiello, A. Sarracino, and M. Zannetti,
J. Stat. Mech. P04003 (2010).

\bibitem{Xtals}
A. Sicilia {\it et al.},
Phys. Rev. Lett. {\bf 101}, 197801 (2008). 

\bibitem{Damien}
D. Bouttes {\it et al.},
%  {\it Dynamical scaling and fragmentation in viscous coarsening: An interrupted in-situ X-ray tomographic study},
  arXiv:1309.1724.

\end{thebibliography}
\end{document}

\maketitle
%%%%%%%%%%%%%%%%%%%%%%%%%%%%%%%%%%%%%%%%%%%%%%%%%%%%%%%%%%%%%%%%%%%%%%%%%%%%%%%

\section{Snapshots}
\begin{figure}[H!]
\begin{center}
\includegraphics[scale=0.375]{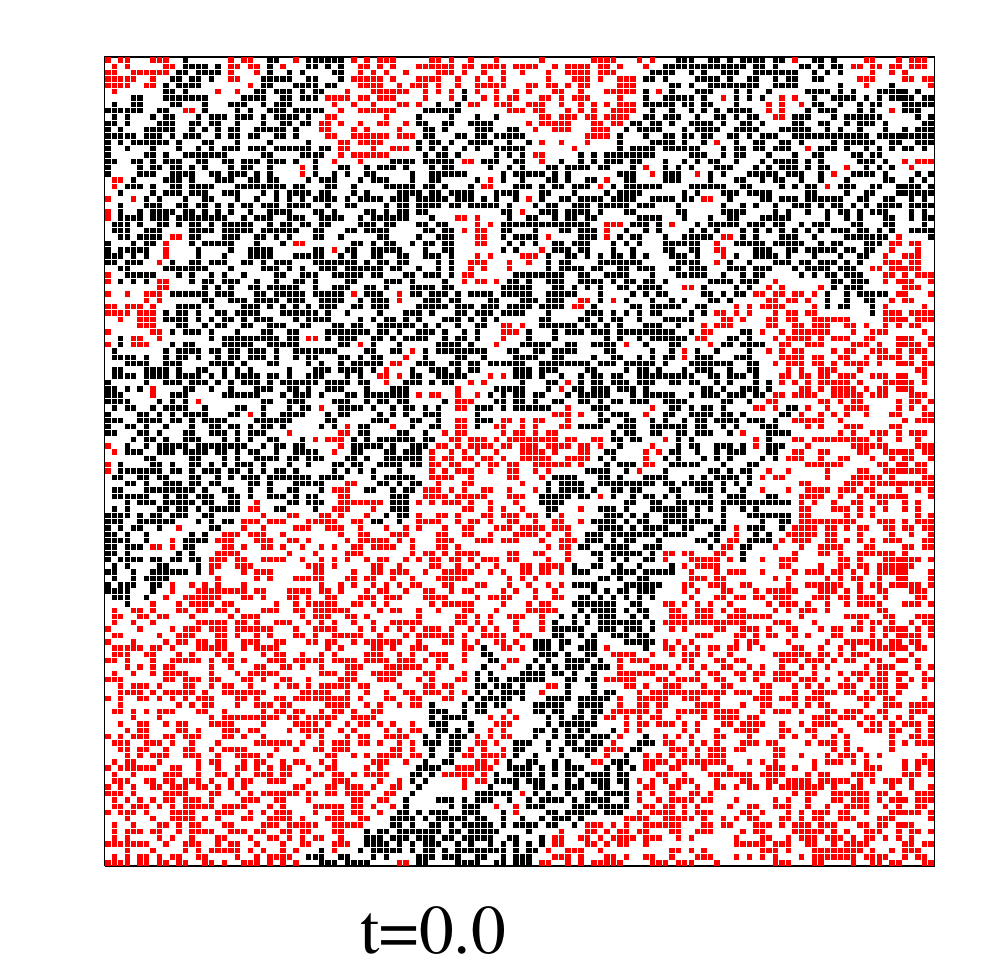}
\includegraphics[scale=0.375]{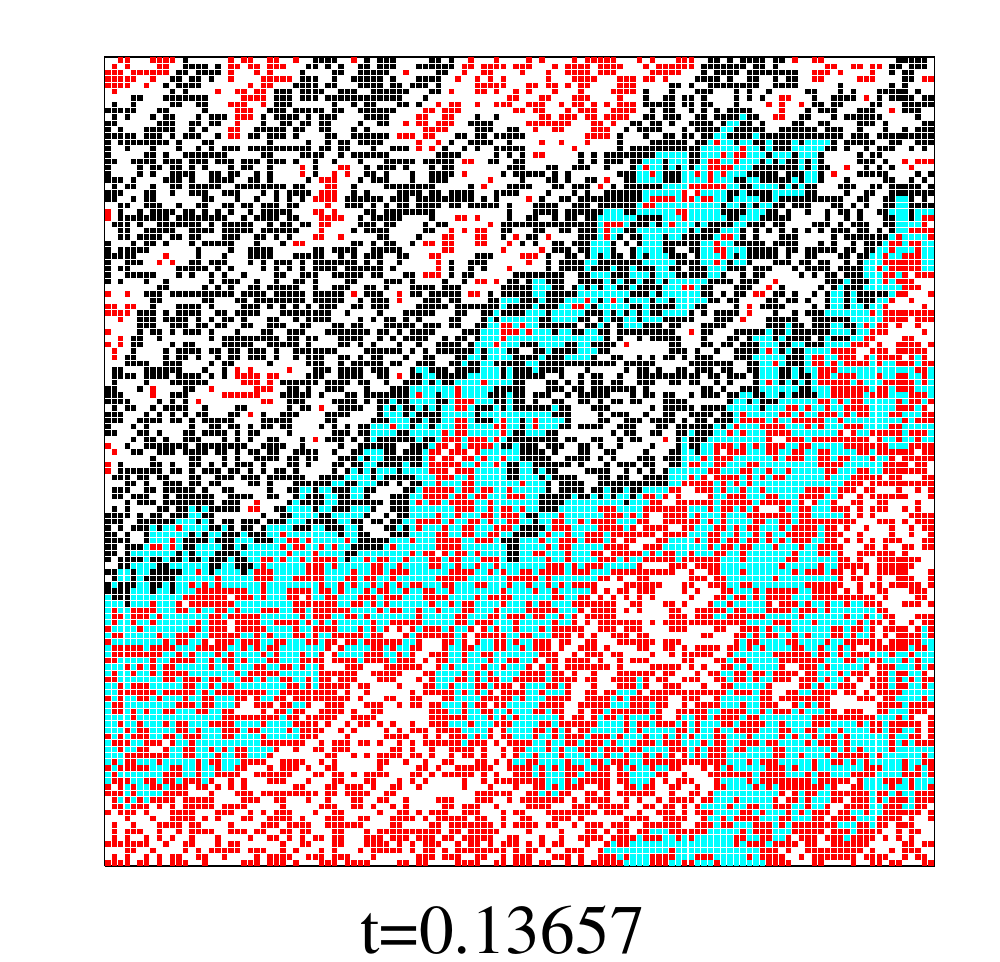}
\includegraphics[scale=0.375]{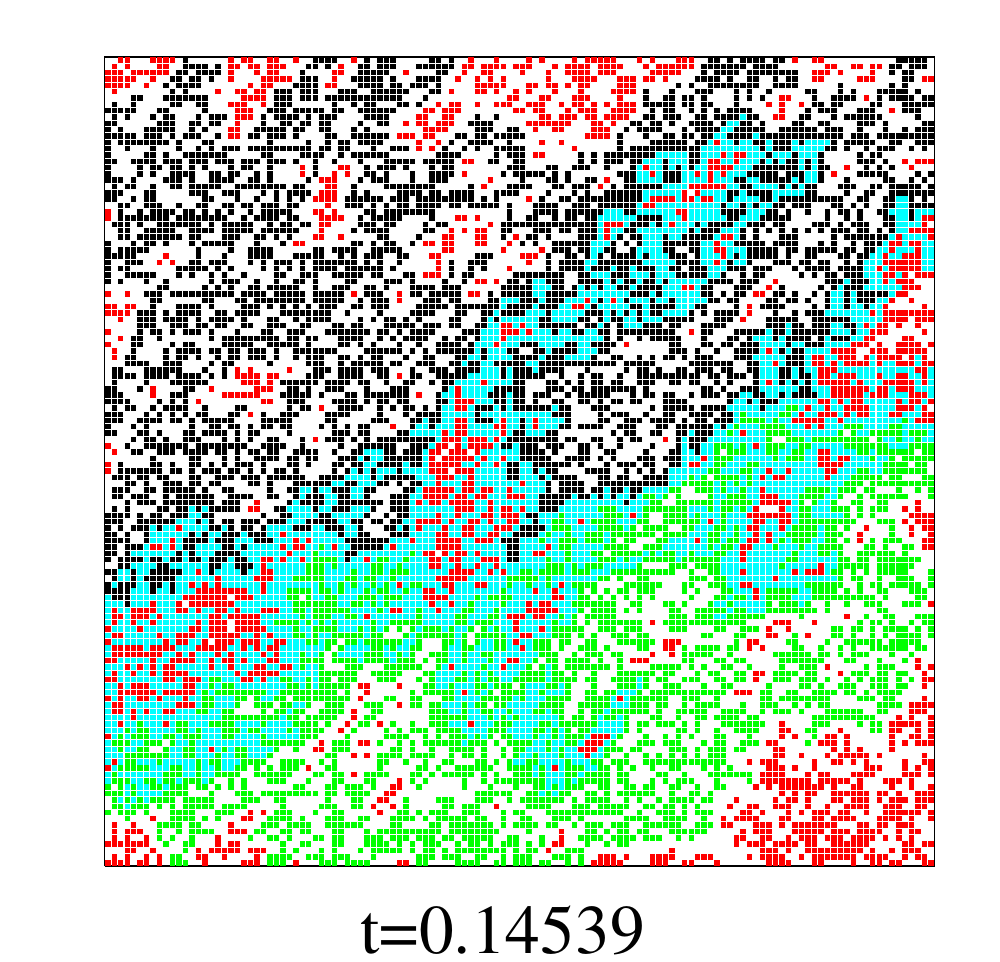}
\includegraphics[scale=0.375]{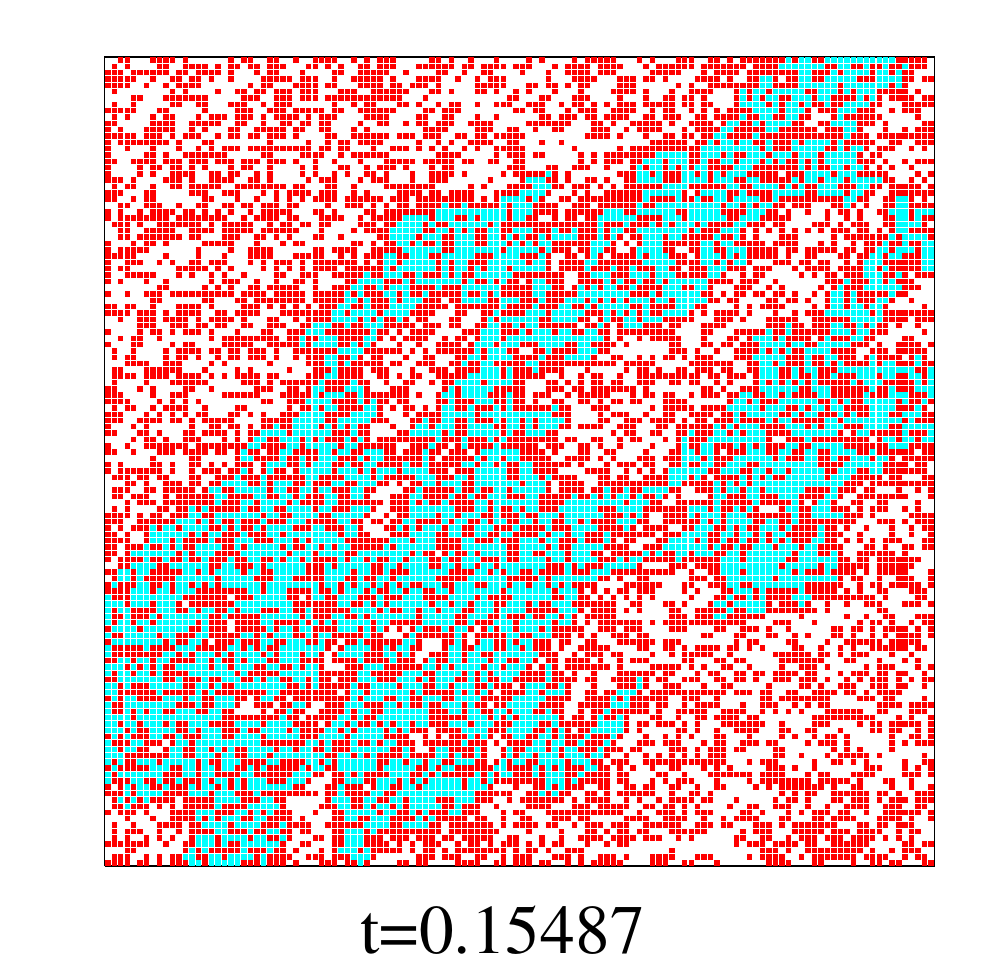}
\includegraphics[scale=0.375]{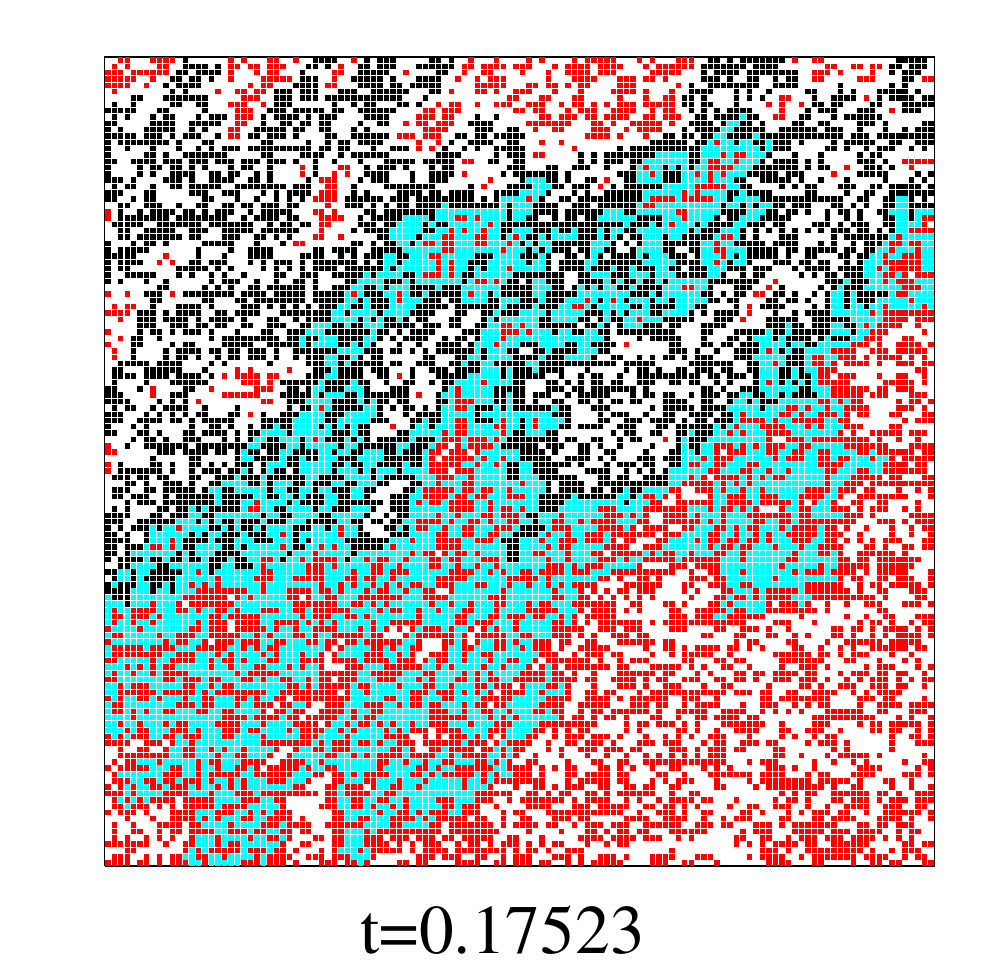}
\includegraphics[scale=0.375]{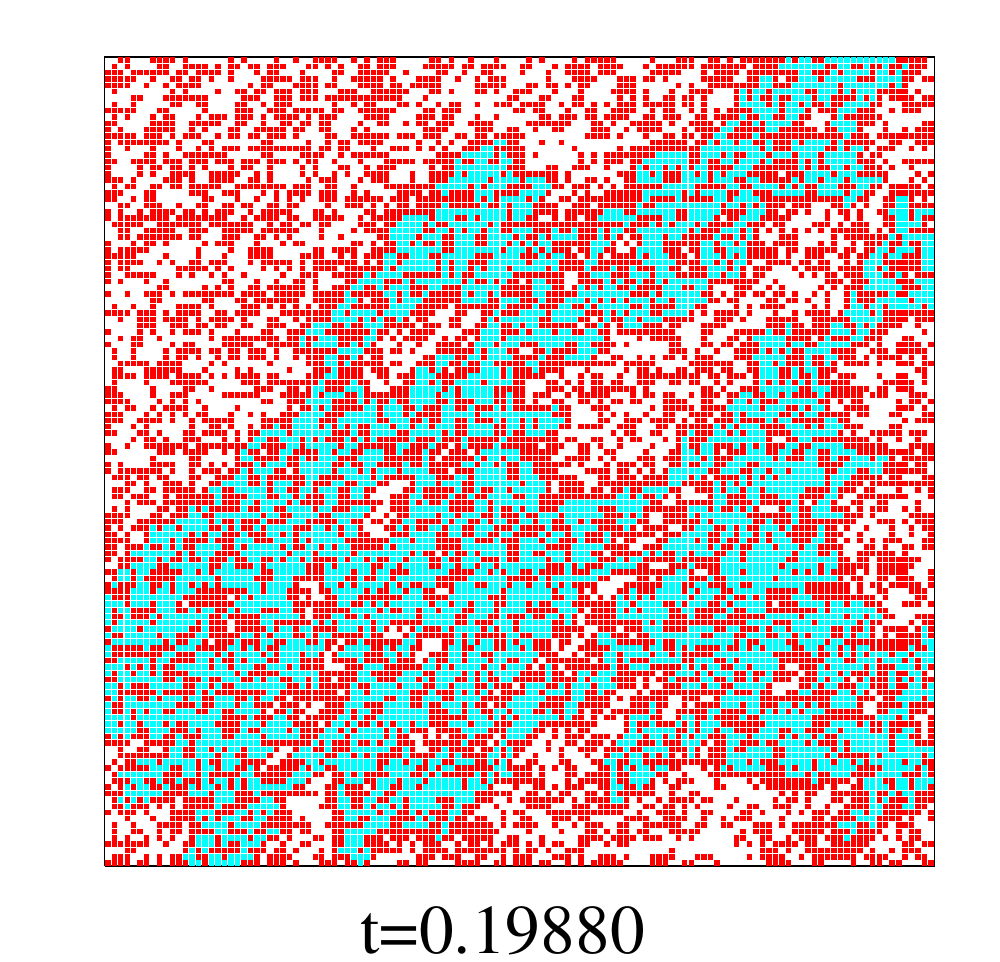}
\includegraphics[scale=0.375]{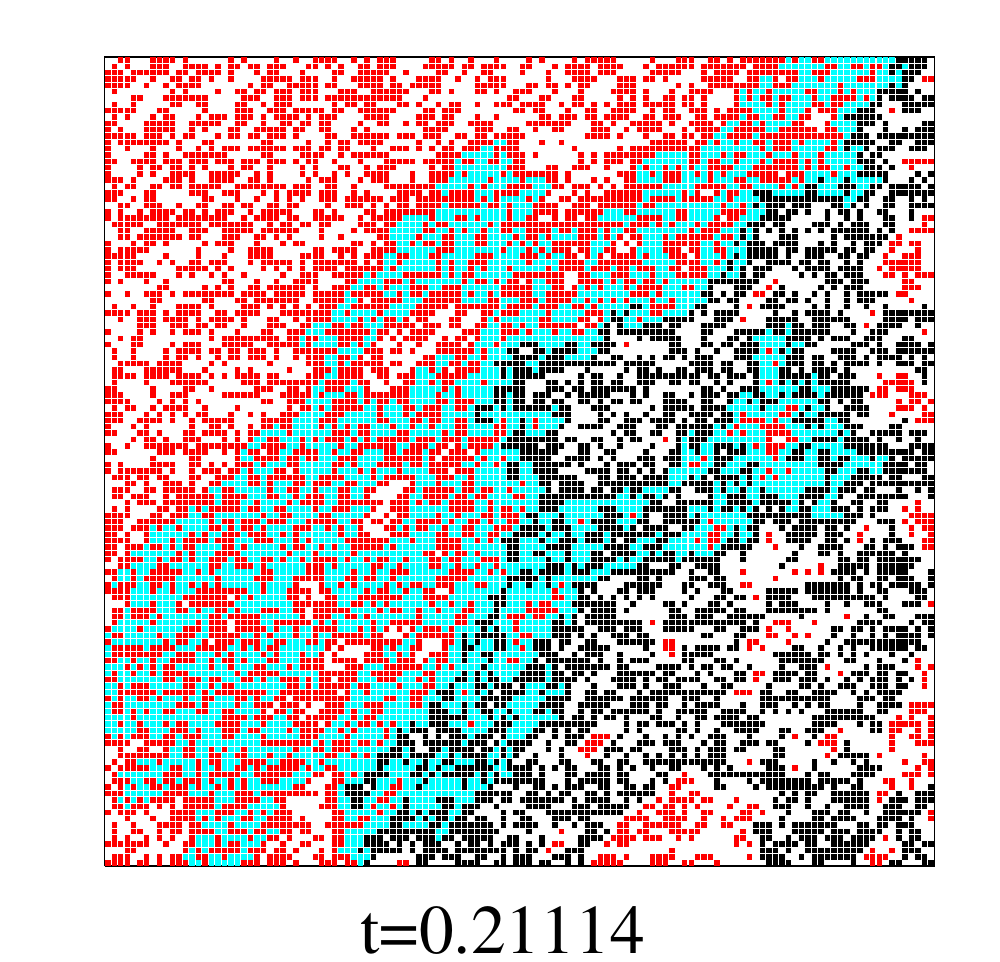}
\includegraphics[scale=0.375]{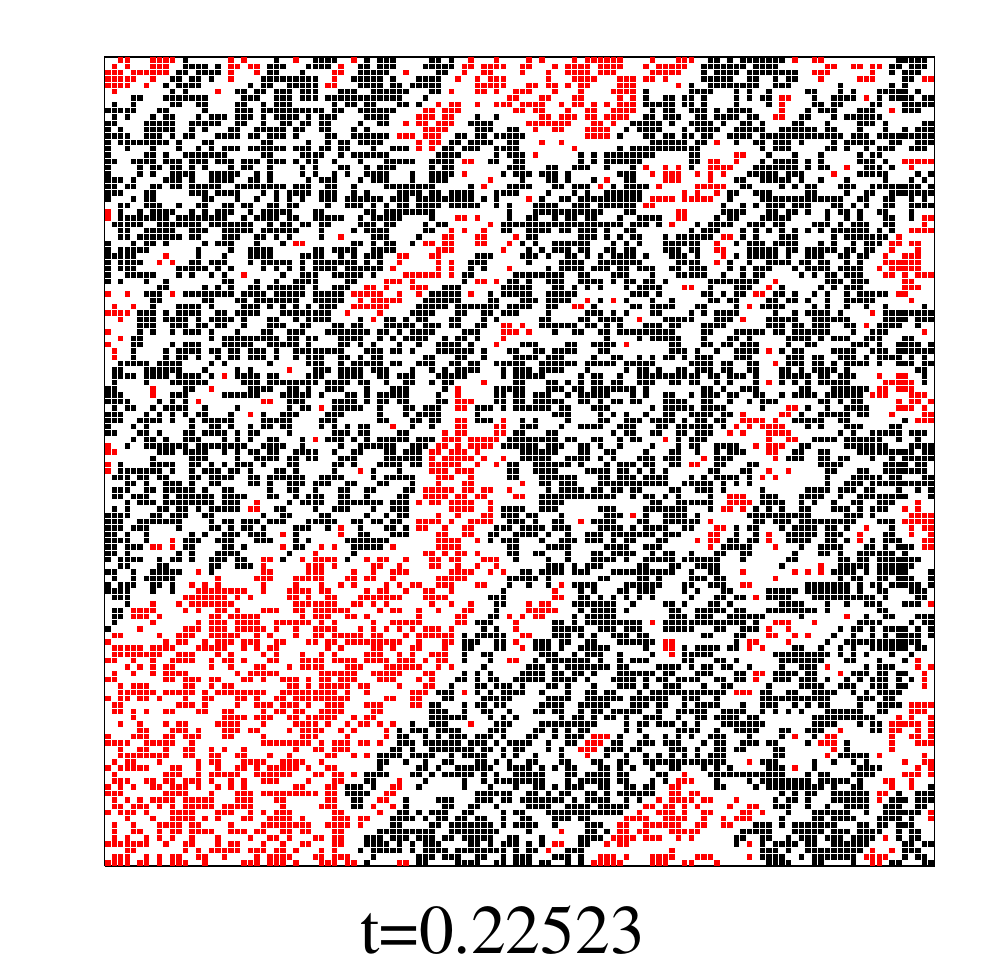}
\includegraphics[scale=0.375]{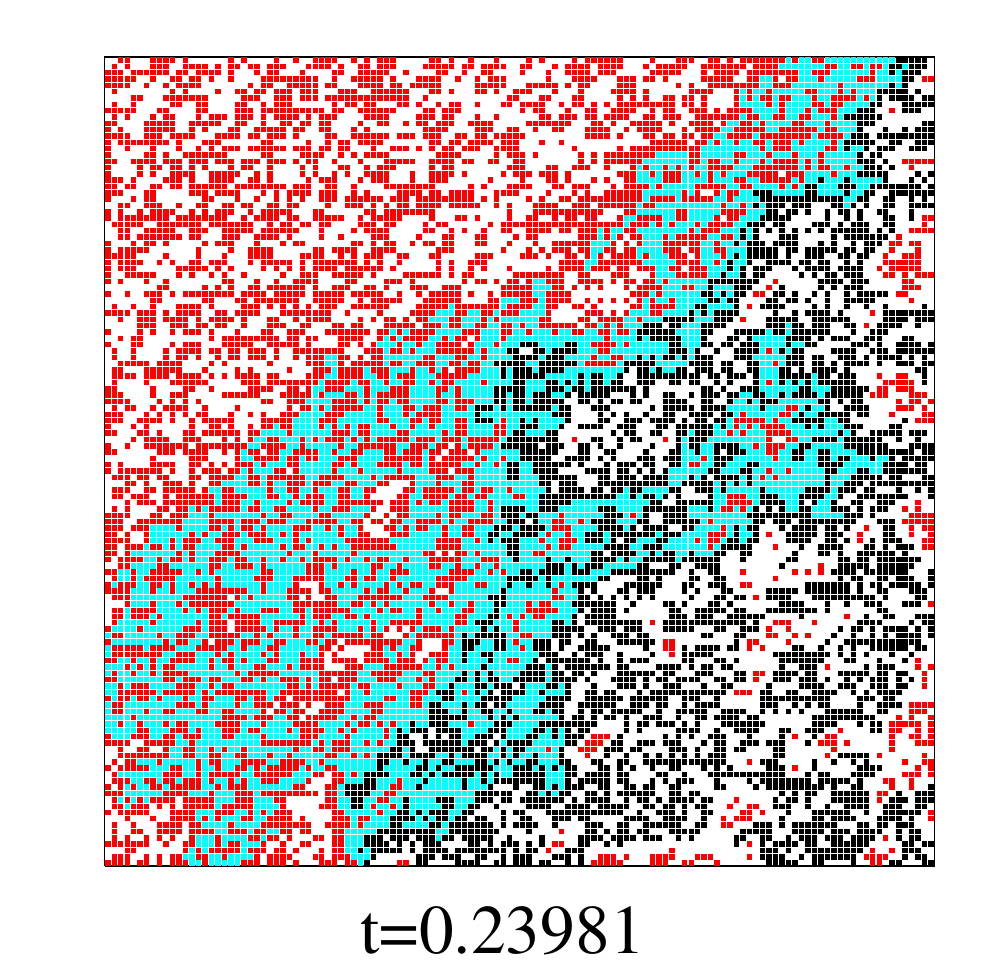}
\includegraphics[scale=0.375]{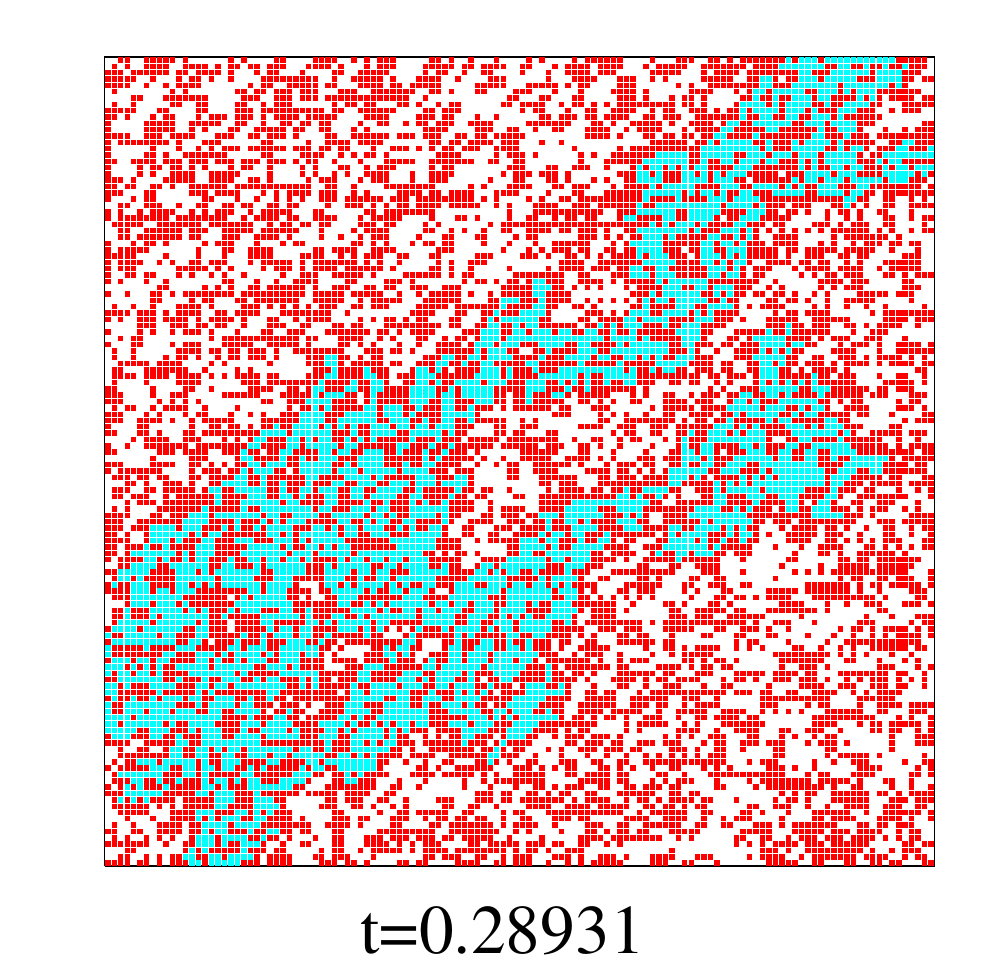}
\includegraphics[scale=0.375]{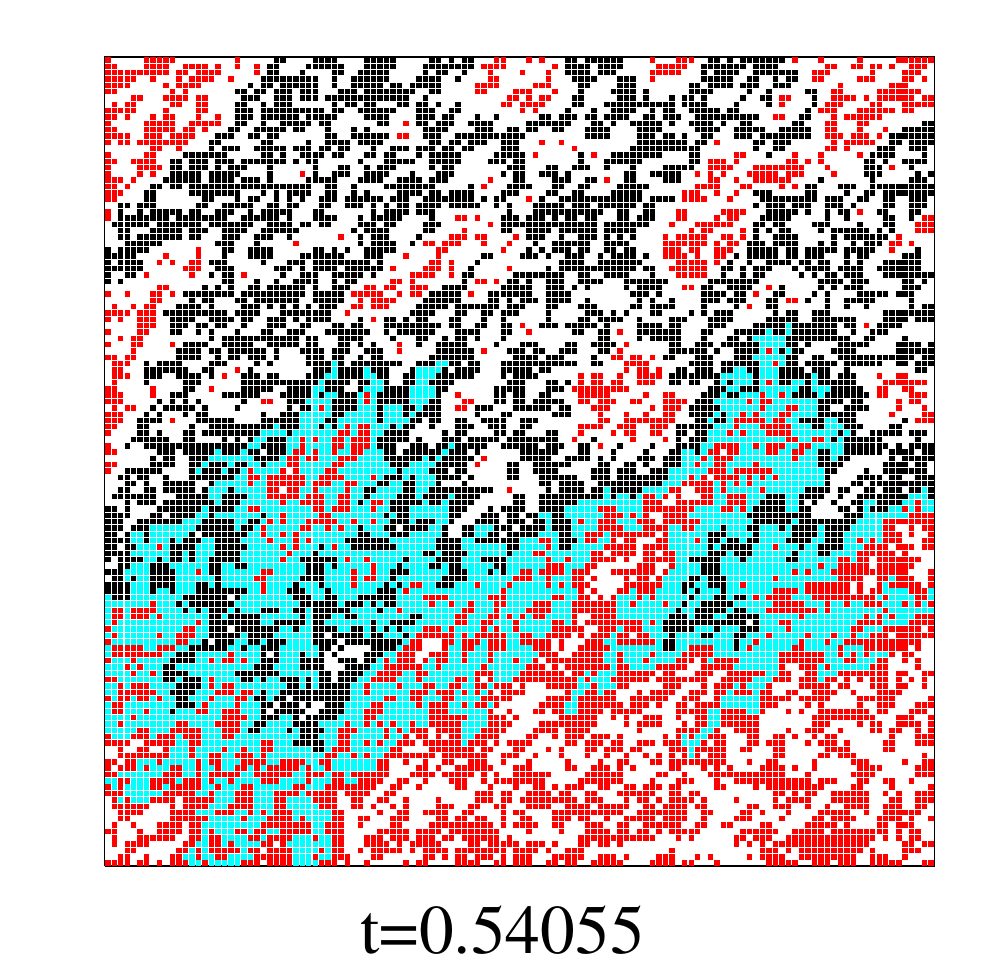}
\includegraphics[scale=0.375]{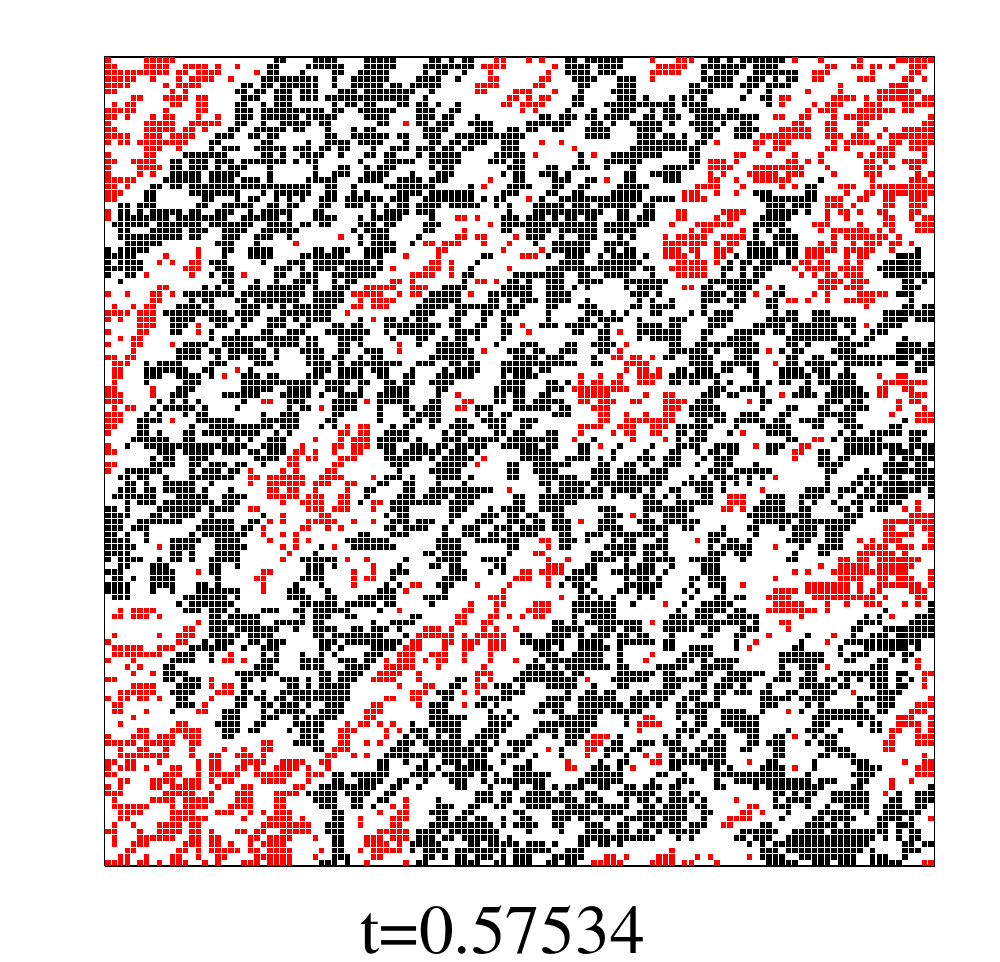}
\includegraphics[scale=0.375]{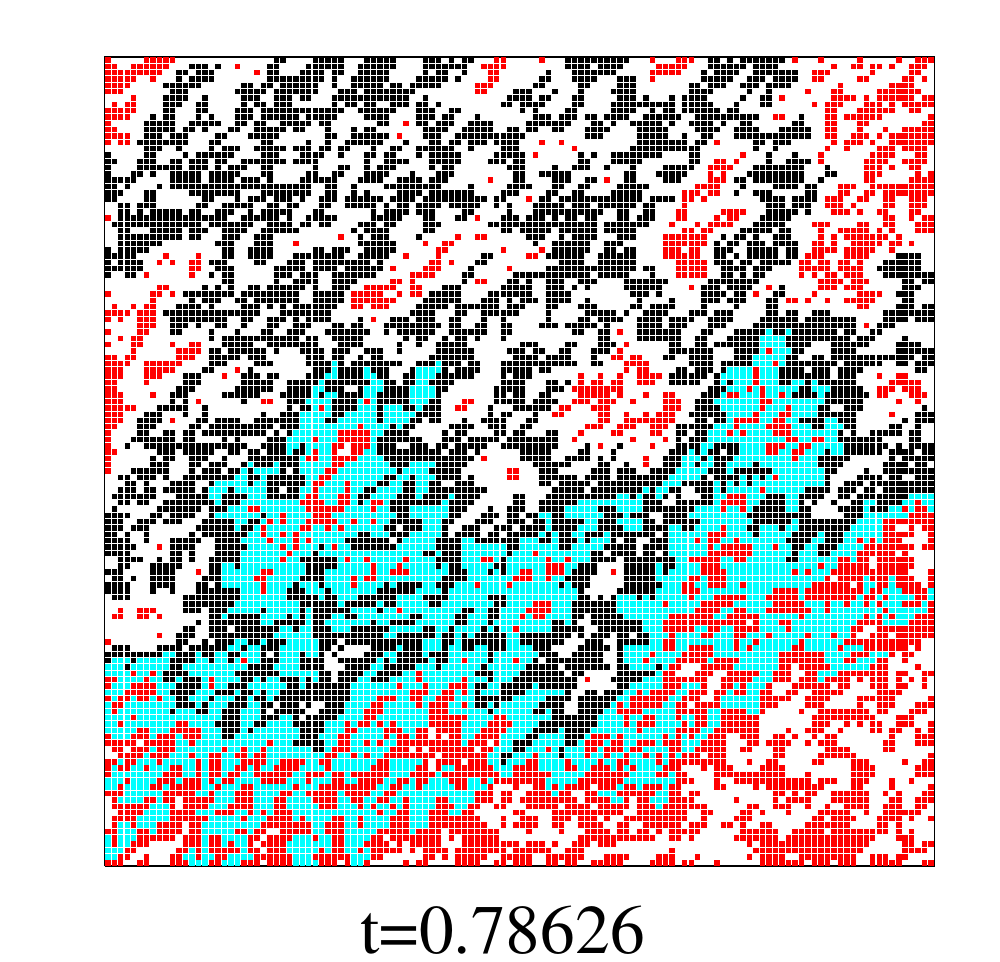}
\includegraphics[scale=0.375]{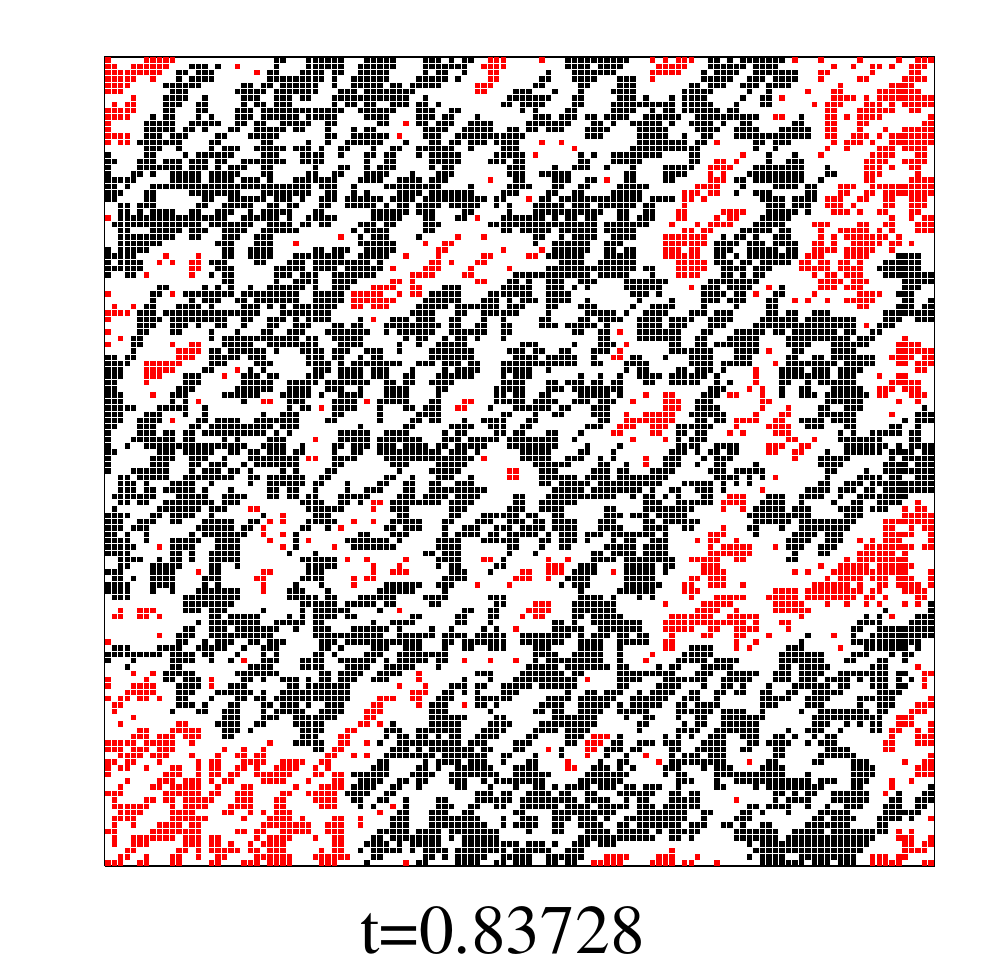}
\includegraphics[scale=0.375]{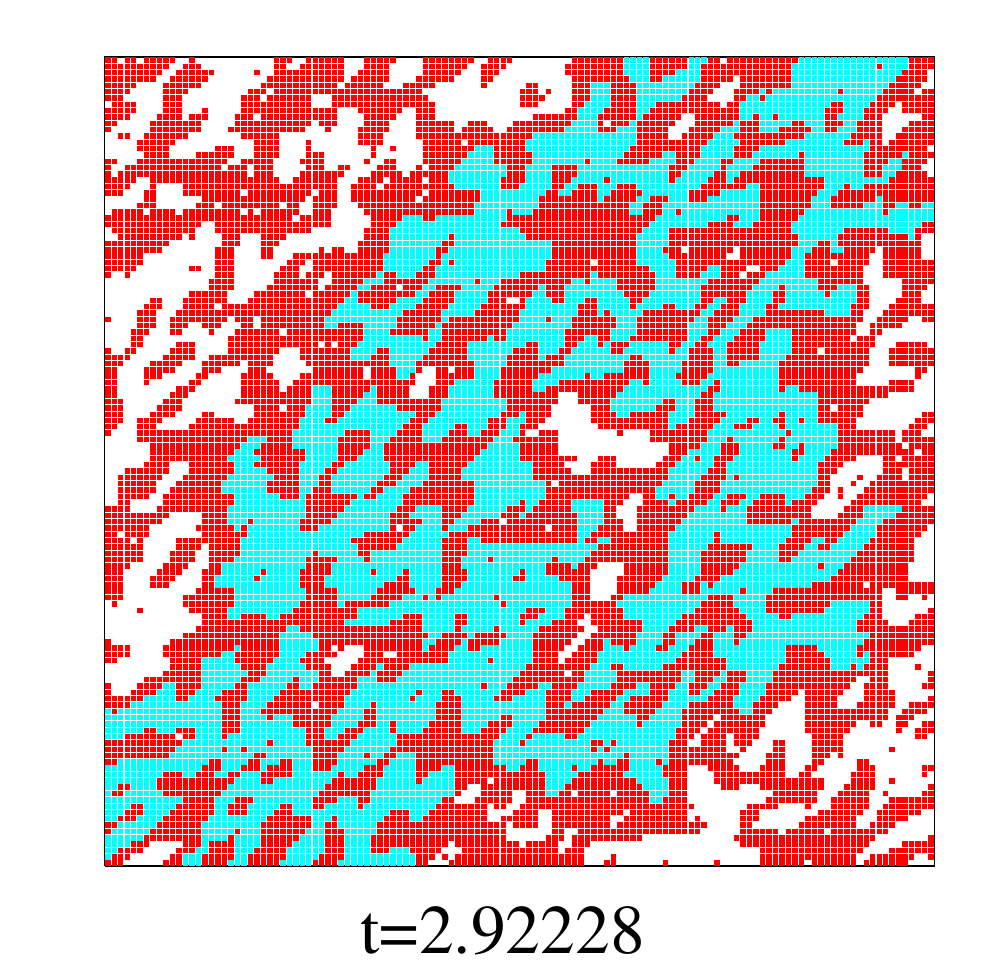}
\includegraphics[scale=0.375]{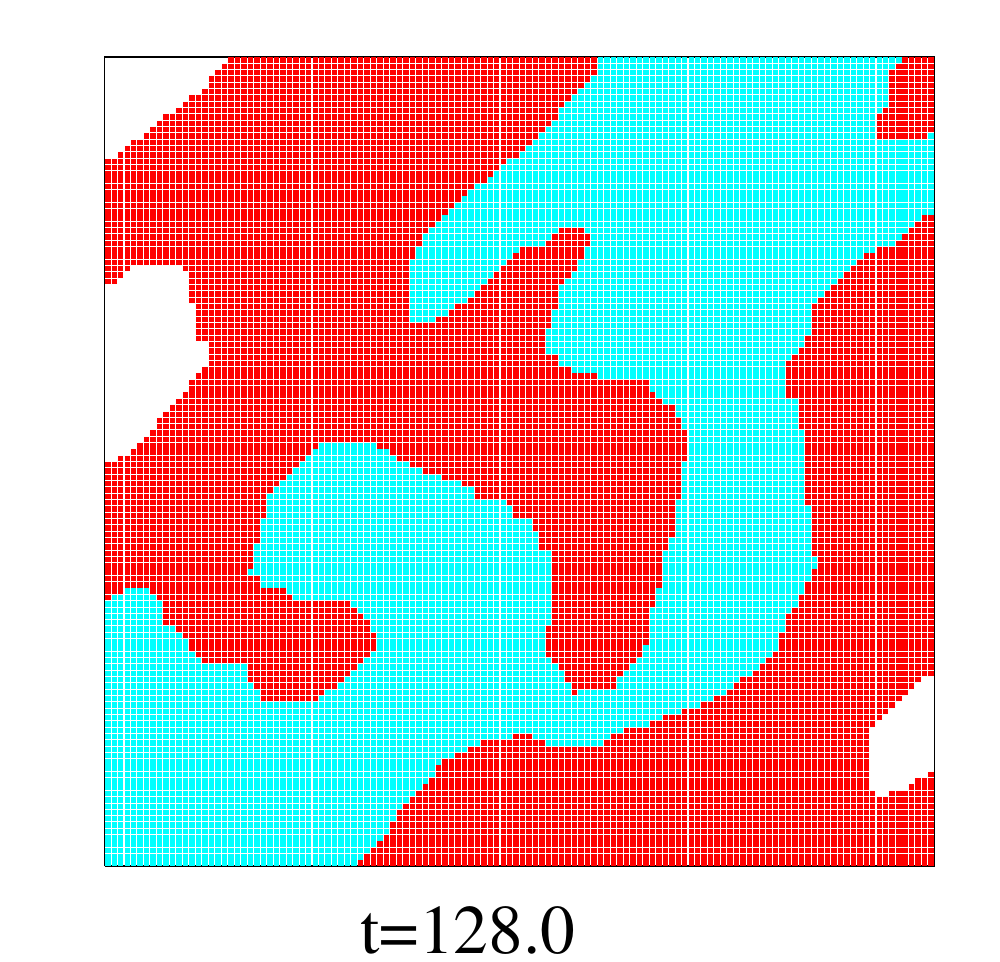}
\vspace{-0em}
\end{center}
\caption{(Color online.)
Snapshots of a $2d$IM on a triangular lattice with $L=128$ 
and FBC. A quench from $T_i \to\infty$ to $T=0$ was performed at $t=0$. 
Spins $S_i=-1 \ (S_i= +1)$ are shown in red (white). 
A percolating cluster of spins $S_i=-1$ ($S_i= +1$) is shown in black or green (clear blue).
}
\label{Percolation}
\end{figure}

We present in Fig.~(\ref{Percolation}) snapshots for a typical evolution on a triangular lattice of linear size $L=128$. The triangular lattice used in the simulation
is built from the square one by adding diagonal bonds in one direction. This explains why the clusters seem to grow
along a diagonal. At the time of the quench $t=0$ a percolating cluster typically exists contrarily to the square
lattice since $p=1/2$ is exactly the site percolation threshold on the triangular lattice. However, this cluster disappears after some time and is
replaced by other percolating clusters similarily to what happens on the square lattice. It is only after many changes  in the
percolative properties of the system that a permanent percolating structure establishes ($t \ge t_p \simeq 2.92$ MCs).
